%% file: main.tex
\documentclass[
    twocolumn,
	prd,
	amssymb,
	preprintnumbers,superscriptaddress,
	nofootinbib]{revtex4-1}

\pdfoutput=1

\usepackage{graphicx}
\usepackage{enumitem}
\usepackage{latexsym}
\usepackage{amsfonts}
\usepackage{amssymb}
\usepackage{xcolor}
\usepackage[export]{adjustbox}
\usepackage{amsmath}
\usepackage{slashed}
\usepackage{dcolumn}
\usepackage{verbatim}
\usepackage{float}
\usepackage{multirow}
\usepackage{xspace}
\usepackage[normalem]{ulem}
\usepackage[
pdfauthor={Djuna Croon}]{hyperref}

\input{universalnewcommands.tex}

\newcommand{\GeV}{\text{GeV}}

\newcommand{\msun}{{\rm M}_\odot}
\newcommand{\rC}{{\rm C}} \newcommand{\rO}{{\rm O}}

\newcommand{\tenx}[1]{ \! \times \! 10^{#1}}
\newcommand{\Eq}[1]{Eq.~(\ref{#1})}

\allowdisplaybreaks

\begin{document}

\title{Missing in Action: New Physics and the Black Hole Mass Gap}
\author{Djuna Croon} \email{dcroon@triumf.ca}
\affiliation{TRIUMF, 4004 Wesbrook Mall, Vancouver, BC V6T 2A3, Canada}
\author{Samuel D.~McDermott}
\email{sammcd00@fnal.gov}
\affiliation{Fermi National Accelerator Laboratory, Batavia, IL USA}
\author{Jeremy Sakstein} \email{sakstein@hawaii.edu}
\affiliation{Department of Physics \& Astronomy, University of Hawai'i, Watanabe Hall, 2505 Correa Road, Honolulu, HI, 96822, USA}
\preprint{FERMILAB-PUB-20-328-T}

\date{\today}

\begin{abstract}
We demonstrate the power of the black hole mass gap as a novel probe of fundamental physics. New light particles that couple to the Standard Model can act as an additional source of energy loss in the cores of population-III stars, dramatically altering their evolution. We investigate the effects of two paradigmatic weakly coupled, low-mass particles, axions and hidden photons, and find that the pulsational pair instability, which causes a substantial amount of mass loss, is suppressed. As a result, it is possible to form black holes of $72\msun$ or heavier, deep inside the black hole mass gap predicted by the Standard Model. The upper edge of the mass gap is raised to $>130{\rm M}_\odot$, implying that heavier black holes, anticipated to be observed after LIGO's sensitivity is upgraded, would also be impacted. In contrast, thermally produced heavy particles would remain in the core, leading to the tantalizing possibility that they drive a new instability akin to the electron-positron pair instability. We investigate this effect analytically and find that stars that avoid the electron-positron pair instability could experience this new instability. We discuss our results in light of current and upcoming gravitational wave interferometer detections of binary black hole mergers. 
\end{abstract}

\maketitle

%%%%%%%%%%%%%%%%%%%%%%%%%
\section{Introduction}
%%%%%%%%%%%%%%%%%%%%%%%%%

Five years after the first detection of gravitational waves from a binary black hole merger \cite{Abbott:2016blz}, the LIGO/Virgo collaboration has detected several dozens of merger events \cite{Abbott:2019ebz}. 
The gravitational waves from these mergers encode the answers to many open questions about the Universe.
The coming decade will see the upgrade of the LIGO/Virgo detectors, as well as the addition of several other gravitational wave experiments. The expected experimental sensitivity will allow for
precision gravitational wave astronomy, including
unprecedented access to the astrophysical population of black holes. This opens up the possibility of using black hole population studies to precisely test competing theories of fundamental physics.

In \cite{Croon:2020ehi}, we proposed the use of the black hole mass gap (BHMG) predicted by stellar structure theory to study physics beyond the Standard Model. In this work, we expand upon that proposal with a more detailed analysis of the mechanisms at work in several models of new physics. As we will show, the anticipated merger event data has the potential to probe several leading models of new, beyond-the-Standard-Model particle physics.  

The existence of the BHMG stems from the \emph{pair-instability} spurred by the production of non-relativistic electron-positron pairs in the cores of massive, low-metallicity population-III stars at the end of their life cycle. The pair production reduces the radiation pressure such that it no longer supports the star against gravitational collapse. The resulting contraction and temperature rise leads to rapid thermonuclear burning of ${}^{16}\rO$, which may release an amount of energy comparable to the star's binding energy. The strength of this explosion, and the subsequent stellar evolution, depend crucially on the mass and metallicity of the parent population-III star. The lighter ($\sim50-90{\rm M}_\odot$ for metallicty $Z\sim 10^{-3}$) progenitors typically undergo a sequence of contractions and explosions in which mass is shed; this has led them to be referred to as pulsational pair-instability supernovae (PPISN). Eventually, these stars return to hydrostatic equilibrium and ultimately core collapse, but the resulting black holes are significantly lighter than they would have been in the absence of the pair instability.
Heavier progenitors ($\gtrsim 90{\rm M}_\odot$ for $Z\sim 10^{-3}$) undergo such a violent explosion that no remnant is left at all, and are referred to as pair-instability supernovae (PISN). The heaviest black hole that can be formed before PPISN losses become significant defines the lower edge of the BHMG. In very heavy progenitors ($\gtrsim240{\rm M}_\odot$ for $Z\sim 10^{-3}$), the pair instability is quenched because some of the energy from the stellar contraction is used to photodisintegrate heavy elements. The lightest black hole formed from this process defines the upper edge of the BHMG.

The stages of the stellar evolution that lead up to the pair instability are particularly volatile. Therefore, small perturbations introduced by new physics may drastically alter the outcome.
We study two distinct scenarios.
Light, bosonic degrees of freedom---frequent protagonists in beyond the standard model theories of particle physics---may be produced copiously and would free-stream out of the star. Using detailed numerical simulations of the evolution of population-III stars from the zero age helium branch (ZAHB), we will explain the effects of these additional losses in detail. In particular, we will show that the resulting mass gap---both the upper and lower edges---is shifted upward in the presence of new physics, implying a new and important science target for mHz-kHz gravitational wave experiments. 
We also study heavy new particles which are produced in thermal equilibrium. These do not free-stream out of the star, but rather reduce the photon pressure directly, as electron-positron pairs do. We take preliminary steps to explore this, by deriving the equation of state for population-III stars including a new component of matter in thermal equilibrium through interactions with the electrons/photons. We find that such new instabilities may indeed be realized in some regions of parameter space. In particular, for new bosons with masses $<m_e$ we find that lower-mass population-III stars which do not encounter the pair-instability could experience this new instability instead. This raises the possibility that thermal production of novel particles could potentially alter the lower edge of the mass gap. A full numerical implementation of this instability is beyond the scope of the present work, but we emphasize that such a treatment would be necessary to determine its consequences and observational signatures.

This paper is organized as follows. In Sec.~\ref{sec:BHMG} we discuss the physics of the pair-instability and the black hole mass gap in more detail, both for the unfamiliar reader's benefit and to gain insight into the effects of novel particle losses. We also discuss potential astrophysical degeneracies that could cause some uncertainty in its precise location. We present an overview of new light bosons, in particular, hidden photons and axions, in Sec.~\ref{sec:DM}. Our numerical code that we use to simulate the effects of new light particles on population-III stars (and to derive the black hole mass gap) is described in Sec.~\ref{sec:MESA}. The main results of our work are presented in Sec.~\ref{sec:DMBHMG}. There, we discuss the effects of novel particle losses on population-III stars, and derive the resultant changes to the location of the black hole mass gap. We explore the possibility that heavier particles produced thermally in the core could lead to a new instability in Sec.~\ref{sec:new_instab}. The derivation of the new instability region is cumbersome, so we present it in Appendix \ref{sec:instab_region} for the interested reader. We discuss our results and conclude in section \ref{sec:concs}.

%---------------------- 

\section{The Black Hole Mass Gap}
\label{sec:BHMG}

\subsection{The Pair-Instability}
The BHMG is the result of ``pair instability'' in the cores of progenitor population-III stars. This instability arises when electron-positron pairs are produced by the thermal plasma: for example, $\gamma\gamma\leftrightarrow e^+e^-$. The threshold energy for this process is $E_{\gamma \gamma} = 2m_e \simeq 1.2 \tenx{10}$K, but $e^+e^-$ pair-production can begin to influence stars whose core temperatures are below $\lesssim10^9$K. This is possible due to the large number of photons in the star: even though the high-energy tail of the Bose-Einstein distribution is exponentially suppressed by $\sim \exp(-10)\simeq 5\tenx{-5}$, this still represents a large number of $e^+e^-$ pairs. Moreover, because these pairs are produced near threshold, they are nearly at rest. As discussed presently, this process robs the star of the pressure support from the relativistic photon plasma, leading to a secular instability.

Massive stars are supported by radiation pressure, so their equation of state (EOS), or first adiabatic index, is given by
\begin{equation}\label{eq:Gamma1}
    \Gamma_1=\left(\frac{\partial P}{\partial \rho}\right)_s\approx \frac{4}{3}.
\end{equation}
As stars with $\Gamma_1<4/3$ are unstable \cite{1968pss..book.....C}, massive stars with $\Gamma_1=4/3$ are on the precipice of an instability, and can be destabilized by a small change in their composition. 
We show the region for which production of $e^+e^-$ pairs causes such an instability in the $T_c$--$\rho_c$ plane in the left panel of Fig.~\ref{fig:BHMG_origin}. 
The boundaries of this region were first derived by reference \cite{1967ApJ...148..803R}; we reproduce their calculation in Appendix \ref{sec:instab_region}.

The shape of the instability region may be understood as follows. At low temperatures (\ie,~the lower left corner of the figure), the Boltzmann suppression is so severe that very few $e^+e^-$ are produced in photon collisions. As the temperature is raised, the process is no longer suppressed, and pairs are produced more readily, yet the majority of the outgoing $e^+e^-$ pairs are non-relativistic as long as $T\lesssim 2m_e$. Such nonrelativistic particles increase the energy density at the core of the star, but not its pressure (\ie,~the non-relativistic pairs do not resist the gravitational compression of the star), so they reduce the EOS. Inevitably, $\Gamma_1$ falls below $4/3$ in some regions of the star. If $\Gamma_1 \leq 4/3$ over a sufficiently large volume of the star, the instability will set in and cause a runaway gravitational collapse. As anticipated above, the lower edge of this region lies as low as $5\tenx8$K. 
At higher temperatures, in particular near and above $m_e$, $e^+e^-$ pairs are produced copiously. However, these particles are relativistic, so they contribute significant pressure to the star and render it stable against gravitational collapse once more.

From Fig.~\ref{fig:BHMG_origin}, we see that the pair instability is removed at high density as well, which requires a different explanation. At high densities, $e^+e^-$ pairs do not appreciably change the EOS of the star because high density stars are supported by the pressure of ions. The ions are non-relativistic; their pressure is a consequence of the ideal gas law $P\propto\rho T$, leading to the familiar EOS of $\Gamma_{1,\,\textrm{ions}}=5/3$. Adding nonrelativistic $e^+e^-$ pairs to a star supported by the pressure of nonrelativistic ions does not lead to an instability.

\subsection{Physical Origin of the BHMG}

\begin{figure*}[t]
    \centering
   { \includegraphics[width=0.49\textwidth]{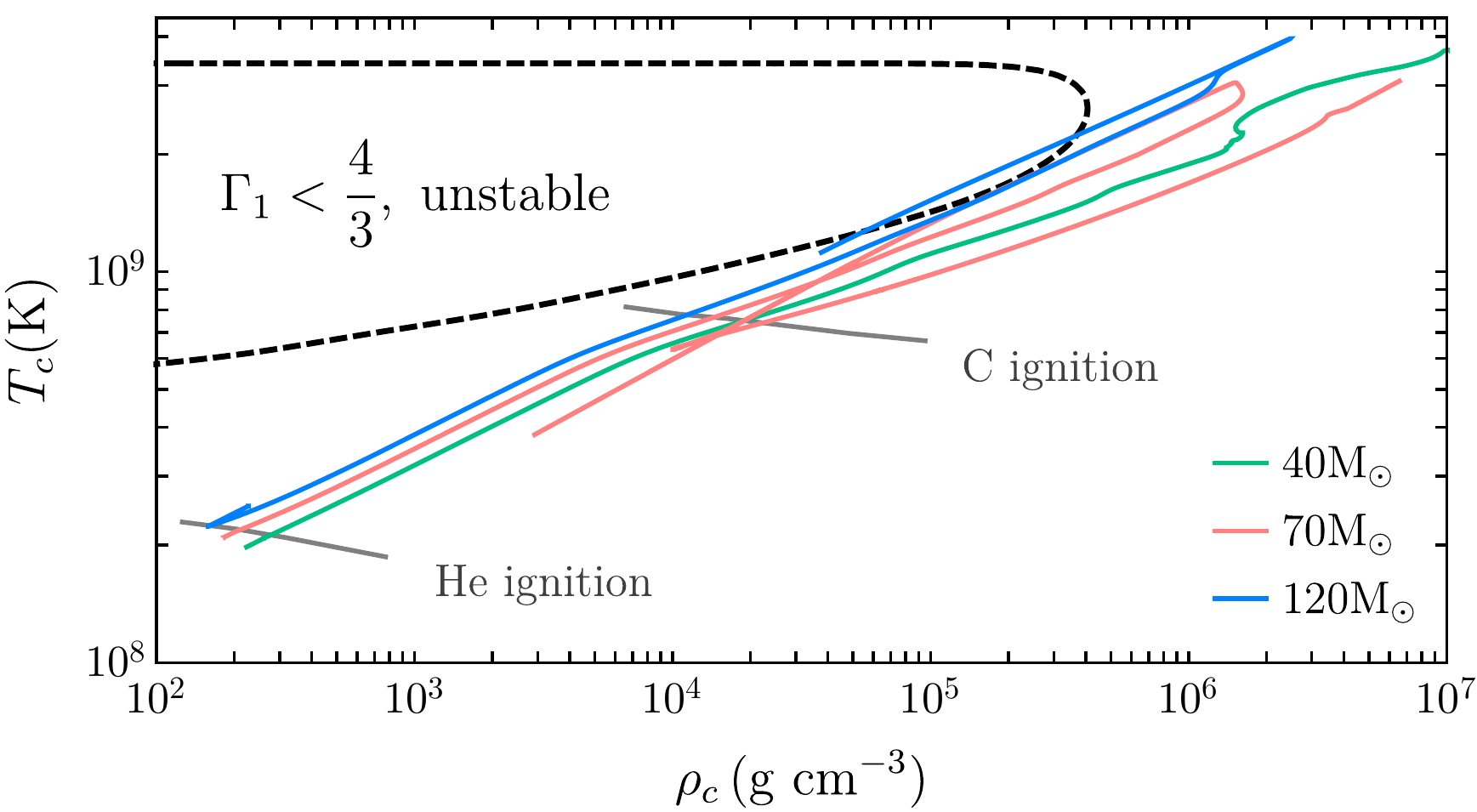}}
  { \includegraphics[width=0.49\textwidth]{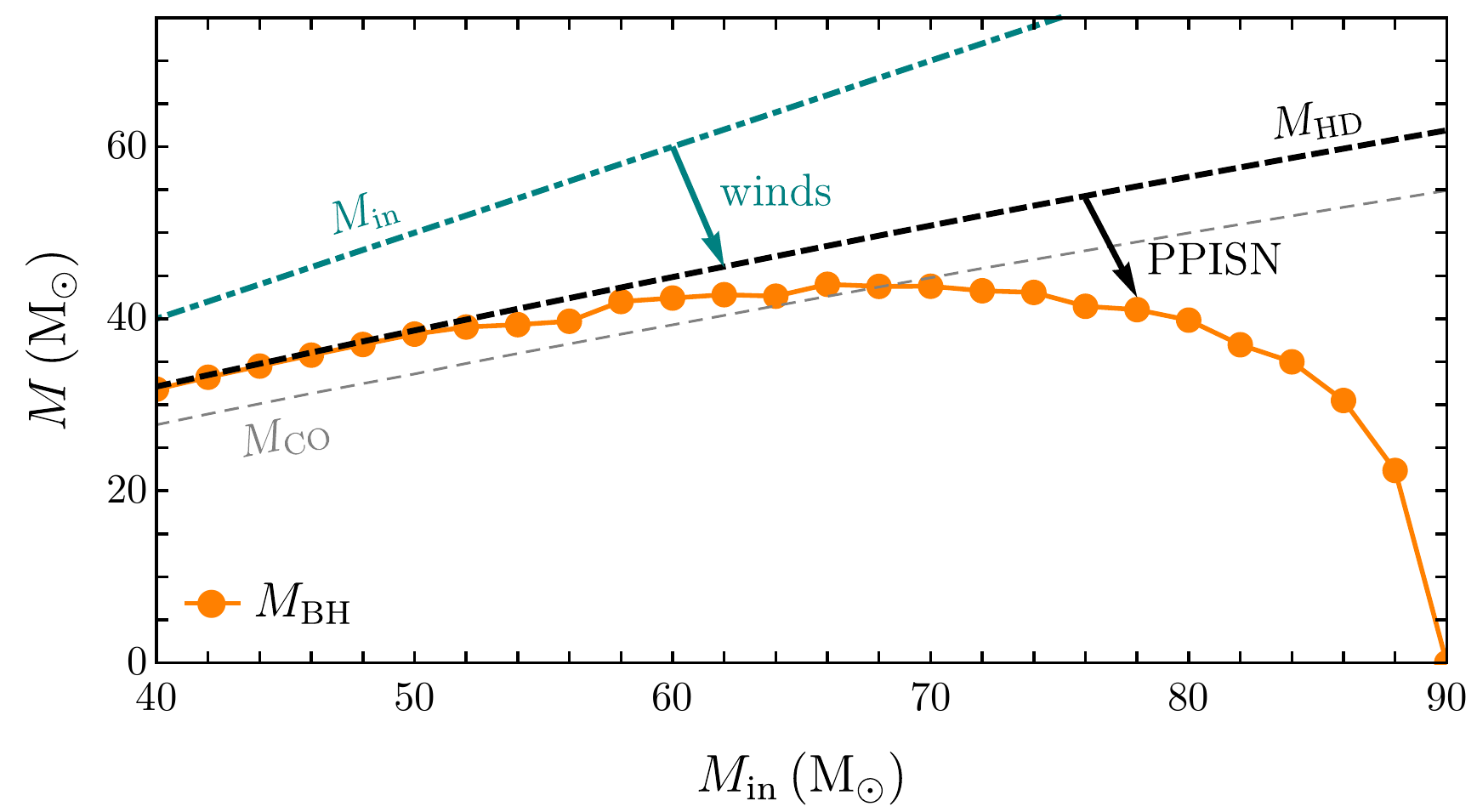}}
    \caption{Demonstration of pair-instability and its effects (adapted from \cite{Croon:2020ehi}).
    \textbf{Left:} 
    The evolution of the central density and temperature of population-III stars of initial metallicity ${\rm Z}_\odot/10$ with initial mass 
    $M_{\rm in}=40{\rm M}_\odot$,
    $70{\rm M}_\odot$, and $120{\rm M}_\odot$ (no new physics is assumed). The black dashed line indicates the region for which the pair-instability occurs. Gray lines indicate the onset of helium and carbon burning. 
    \textbf{Right:} Various masses as a function of initial stellar mass $M_{\rm in}$ for population-III stars of initial metallicity ${\rm Z}_\odot/10$. The teal dot-dashed line shows the initial mass, the black dashed line shows the entire mass of the star at helium depletion, and the gray dashed line shows the CO core mass at helium depletion. The orange points correspond to final black hole masses for individual stellar models. The PISN occurs at higher masses, leaving no compact object in the final state.
    }
    \label{fig:BHMG_origin}
\end{figure*}

A star's final fate is intimately linked with its initial mass and composition \cite{2002ApJ...567..532H}. A star with an initial mass $\lesssim 8\msun$ will end its life as a white dwarf, while stars with initial masses between $8\msun\lesssim M_i \lesssim 25\msun$ (where the upper value depends on metallicity, $Z$) will end their lives as neutron stars. Above this mass, the core collapse process that is triggered when the star runs out of nuclear fuel is not completely slowed by the stiffening of the nuclear equation of state. Instead, these stars either become black holes or experience such a violent explosion that they unbind entirely. The BHMG is a range of black hole masses which stellar structure theory predicts will be unpopulated: no stars exist whose final fate is a black hole with mass in the mass gap. The gap lies between three distinct final fates for a massive star: direct core collapse, pulsational pair instability supernova (PPISN), and pair instability supernova (PISN).

Direct core collapse is the outcome for a star with initial mass in the range $25\msun \lesssim M_{\rm in} \lesssim 50\msun$ (where the bracketing values are $Z$ dependent). Such stars never experience the pair instability discussed above, but instead experience a gravitational instability and collapse directly to a black hole after they establish a sufficiently heavy iron core mass and run out of combustible fuel. This would be the fate of all massive stars if not for the pair instability. The pair instability triggers contraction when the star still has a significant fraction of combustible material. Thus, after the collapse is initiated, and the density is increased throughout the stellar volume, violent fusion reactions can take place, with different effect depending on the ratio of ${}^{12}$\rC ~ to ${}^{16}$\rO. For the range of initial stellar masses $50\msun \lesssim M_{\rm in} \lesssim 90\msun$ (for $Z\sim 10^{-3}$), the star will expand again after the initial contraction induces thermonuclear burning, in a cycle known as a pulsation. The pulsation ejects loosely bound material from the outer volume of the star; this can happen once or multiple times, and the star can lose a small or large fraction of its mass in each pulsation. If the star experiences one or more pulsations but maintains a gravitationally bound core, the star is said to undergo the PPISN. At higher initial mass $90\msun \lesssim M_{\rm in} \lesssim 240\msun$ (again, for $Z\sim10^{-3}$), the initial contraction results in such an intense period of explosive oxygen burning that the star unbinds entirely. This explosion with no black hole remnant defines the PISN. Finally, for the highest initial masses, $M_{\rm in} \gtrsim 240\msun$ (and $Z=10^{-3}$), the freely falling stellar material heats up to such a degree on its inward journey that the nuclei photodisintegrate. This prevents the PISN and instead leads to direct collapse.

We illustrate these processes by charting the evolution of population-III star progenitors from the onset of helium burning through to core collapse in Figure~\ref{fig:BHMG_origin}. The left panel shows three stellar tracks (with metallicity $Z={\rm Z}_\odot/10=0.00142$) in their central density $\rho_c$ and temperature $T_c$ as they progress from the zero age helium branch (ZAHB) onward. In the beginning of this evolution, all three stars burn helium in their core primarily via the triple-alpha process $3 \alpha \rightarrow {}^{12}\rC+2\gamma$, as well as in the (subdominant) carbon alpha capture process ${}^{12}\rC(\alpha,\gamma)^{16}\rO$. This leads to an increase in both $\rho_c$ and $T_c$; importantly, as we discuss later, this also builds up oxygen throughout the stellar volume. During helium burning, the stars are also continually losing mass due to stellar winds (we give our prescription for wind loss rates in Sec.~\ref{sec:MESA}). In this phase of life, their evolution proceeds essentially in parallel. After the initiation of carbon burning, however, the tracks diverge. The green track in this panel corresponds to a $40\msun$ star. This star avoids the pair instability and continues to burn fuel to increasingly high densities, before exhausting its fuel and undergoing direct collapse to a black hole. The only mass it has lost is due to winds. The pink track follows the evolution of a $70\msun$ star. This star experiences pulsations that cause it to lose a substantial fraction of its mass. However, these pulsations of the PPISN are not strong enough to unbind the star completely, and it eventually relaxes to hydrostatic equilibrium before collapsing to form a black hole. The third track depicts a $120\msun$ star. This star experiences a PISN that unbinds the star entirely.

We summarize the impact of winds and pulsations on stars of metallicity ${\rm Z}_\odot/10$ in the right panel of Fig.~\ref{fig:BHMG_origin}. As a function of the initial stellar mass, we show: the initial mass, $M_{\rm in}$; the total mass at the time of helium depletion, $M_{\rm HD}$; the carbon-oxygen core mass at the time of helium depletion, $M_{\rm CO}$; and the final black hole mass, $M_{\rm BH}$ (see section \ref{sec:MESA} for the precise definitions of these). In the absence of winds and pulsations, every final black hole mass would be equal to the initial mass. Winds are important for stars of all masses, however, and the stellar mass at the time of helium depletion will be appreciably lower than the initial for all stars we study. This translates to a final black hole mass that is constrained to have $M_{\rm BH} \leq M_{\rm HD} < M_{\rm in}$. Stars with $M_{\rm in} \lesssim 50\msun$ avoid the pair instability entirely, such that winds are the only source of mass loss, and we find $M_{\rm BH} \simeq M_{\rm HD}$. Stars with a somewhat larger ZAHB mass, in the range of $50\msun \leq M_{\rm i} \leq 90 \msun$, lose mass to pulsations due to the PPISN but ultimately collapse to form a black hole. This is shown by the fact that $M_{\rm BH}$ is nonzero but is strictly less than $M_{\rm HD}$ in this range. Finally, stars with an initial mass $M_{\rm in} \geq 90\msun$ experience the PISN, and $M_{\rm BH}$ goes to zero. At large enough $M_{\rm in}$ (for the metallicity studied here, this happens for stars with $M_{\rm i} \geq 240\msun$), the PISN is thwarted due to energy losses from the photodisintegration of heavy elements and black holes can once again form, with masses $M_{\rm BH} \geq 122\msun$. 

The BHMG is the range of masses in between the heaviest PPISN and the lightest star whose infall disrupts the PISN. As discussed presently, there are Standard Model uncertainties on these values, as well as stochastic parameters such as composition, rotation, and binarity, that are expected to blur the boundaries somewhat \cite{Marchant:2018kun,Farmer:2019jed,Farmer:2020xne}. The impact on the BHMG of new energy loss mechanisms, which we suggest can come from light weakly coupled particles like axions and hidden photons, will be the focus of Sec.~\ref{sec:DM} and beyond.

%%%%%%%%%%%%%%%%%%%%%%%%%
\subsection{Known physics dependence of the BHMG}
%%%%%%%%%%%%%%%%%%%%%%%%%
The precise location of the black hole mass gap is sensitive to several processes that affect stellar evolution. There is both {\it inherent spread} due to stochastic random variables that take on different values in different stars (\eg,~wind loss and metallicity), depend on environment (\eg,~binarity), or for which we have limited prescriptions in our stellar modelling (\eg,~rotation and mixing); and there is {\it uncertainty} due to measurement uncertainties in the input physics (\eg,~theoretical nuclear reaction rates and neutrino loss rates). The effects of these have recently been investigated by \cite{Farmer:2019jed, Farmer:2020xne}. We briefly discuss some important contributions here to highlight possible degeneracies with new physics, and to gain some insight into how novel particle losses affect the BHMG: 
\begin{itemize}
    \item {\bf Metallicity:} Metallicity affects $M_{\rm BH}$ because the mass lost to winds during core helium burning scales as $Z^{0.85}$ \cite{Vink:2001cg,Vink:2005zf,Brott:2011ni}. Lower metallicity stars lose a smaller portion of their mass, and thus are able to form heavier black holes. Interestingly, \cite{Farmer:2019jed} found that the lower edge of the mass gap is relatively robust against changes in metallicity, shifting only by $\sim3 \msun$ over the range $10^{-5}<Z<3\times10^{-3}$ relevant for the population-III stars that are the progenitors of the black holes observed by LIGO/Virgo. Lower metallicity stars correspond to higher-mass black holes because less mass is lost to winds before PPISN commences, and since we are primarily interested in the lower edge of the BHMG we show results for $Z=10^{-5}$ in what follows. When disucssing the upper edge we will also show results for $Z={\rm Z}_\odot/10$.
    \item {\bf Wind loss:} The wind loss efficiency parameter $\eta$ is subject to some variation due to the effects of clumping \cite{Yoon:2010at}. Larger values of $\eta$ result in more mass loss and therefore lighter black holes. Again, \cite{Farmer:2019jed} found that the location of the lower edge of the black hole mass gap is relatively robust to variations in the wind loss prescription, showing differences of at most $3 \msun$ for three different mass loss prescriptions and values of the clumpiness parameter $\eta = \{0.1,1\}$. We use the prescription of \cite{Yoon:2010at} with $\eta = 0.1$ for our results.
    \item {\bf Nuclear physics:} In this work we use default MESA rates (a combination of central values from the {\tt NACRE} \cite{Angulo:1999zz} and {\tt REACLIB} \cite{2010ApJS..189..240C} databases) for all reactions in order to allow for a direct comparison with previous works. However, it is known that the physics of the pair-instability renders the final black hole mass very sensitive to the amount of ${}^{12}$C present when pulsations begin \cite{Farmer:2019jed,Farmer:2020xne}. The reason this affects the BHMG is that inward contractions raise the core temperature during the pulsations, igniting residual carbon, which in turn explosively ignites the residual oxygen, resulting in the outward shock responsible for the most substantial mass loss. The most important rate for determining the amount of carbon is the rate for ${}^{12}\rC(\alpha,\gamma)^{16}\rO$, which, as mentioned above, converts carbon to oxygen during core helium burning \cite{Farmer:2019jed}. Decreasing this rate results in a larger proportion of carbon in the CO core, and results in a less explosive star. Recently, \cite{Farmer:2020xne} explored the impacts of the ${}^{12}\rC(\alpha,\gamma)^{16}\rO$ rate on the BHMG in more detail, finding that $+(-)1\sigma$ variations in the rate affected the lower boundary of the mass gap by $\simeq -1(+5)\msun$, but that $+(-)3\sigma$ variations in this rate led to a change in the lower boundary of the BHMG of $\simeq -2(+50)\msun$. However, the spread of rates used for the main results of \cite{Farmer:2020xne} reflect error bars from \cite{Sallaska:2013xqa}, which ultimately derive from \cite{2002ApJ...567..643K}. This spread is substantially larger than in the most recent compilation \cite{deBoer:2017ldl}. Thus, it is highly unlikely that the uncertainties profiled over in \cite{Farmer:2020xne} are compatible with current, high-precision data \cite{nuclearconvo}. When using errors derived from \cite{deBoer:2017ldl} instead, \cite{Farmer:2020xne} find that the mass gap varies only by $\simeq -4(+0)\msun$. Thus, we expect that the uncertainty due to ${}^{12}\rC(\alpha,\gamma)^{16}\rO$ is not substantially more pronounced than other uncertainties. We will assume the lower edge of the mass gap has error bars $\simeq -4(+0)\msun$ due to this reaction \cite{Farmer:2020xne,deBoer:2017ldl}.
    \item {\bf Other physics:} Reference \cite{Farmer:2019jed} found that the effects of other uncertain physics such as mixing and electroweak uncertainties on the neutrino loss rate have only a minor impact, changing the location of the lower edge of the mass gap by $1\msun$ or less.
\end{itemize}
It is beyond the scope of the preliminary exploratory work we present here to fully investigate these, but we believe that exploring and accounting for these degeneracies will inevitably be a critical step before conclusively interpreting the BHMG as a sign of (or a constriction on) new physics.

%%%%%%%%%%%%%%%%%%%%%%%%%
\section{Light particle emission}
%%%%%%%%%%%%%%%%%%%%%%%%%
\label{sec:DM}

In this section, we describe how light, non-Standard Model particles can be emitted from massive stars.

\subsection{Electrophilic Axion}

We begin our study with an ``electrophilic axion'', the same model we discussed in \cite{Croon:2020ehi}. The Lagrangian involving the axion-like particle (henceforward axion) $a$ is
\begin{align} \label{Lae}
    \cL &= \cL_{\rm SM} +\frac1{2\Lambda} \partial_\mu a \bar\psi_e \gamma^\mu \gamma_5 \psi_e a - \frac12 m_a^2 a^2
    \\&\to \cL_{\rm SM} - i g_{ae} a \bar\psi_e \gamma_5 \psi_e a - \frac12 m_a^2 a^2.\nonumber
\end{align}
In \Eq{Lae}, $g_{ae}$ is a dimensionless coupling which arises from the shift-symmetric derivative interaction in the first line. This interaction is suppressed by a mass scale $\Lambda$, which should be large to avoid introducing new light particles coupled to the electron, so we expect $g_{ae} = m_e/\Lambda$ is a small number. In \Eq{Lae}, $\psi_e$ is the electron Dirac field, and $m_a$ is the axion mass, which we assume for now is much smaller than the temperature of the star.
For ease of presenting results, we will work with the quantity $\alpha_{26} \equiv 10^{26} \alpha_{ae} \equiv 10^{26} g^2_{ae}/4\pi$. As we discuss in more detail below, stars are sensitive to couplings $\alpha_{26}\sim \cO(1-100)$.

We will mostly be interested in temperatures of order $10^8 - 10^9$K; in this range, the electrons in the star are non-relativistic but the axions are effectively massless. In low-metallicity objects like the population-III stars of interest to us, the primary effects are from semi-Compton and bremsstrahlung processes\footnote{The ``A'' processes (axio-recombination and -deexcitation) depend sensitively on metallicity \cite{Redondo:2013wwa}, and are subdominant in the systems of interest for this study.}. The specific energy loss rate due to axionic semi-Compton scattering, $e+\gamma \rightarrow e+a$, is given by \cite{Raffelt:1994ry},
\begin{eqnarray}\label{eq:Q_SC}
    \mathcal Q_{\rm sC} \! = \! \frac{160 \,\zeta_6 \alpha_{\rm EM} \alpha_{ae}}{\pi}  \, \frac{Y_e T^6F_{\rm deg}}{m_N m_e^4} \!\simeq
    33\alpha_{26}Y_eT_8^6F_{\rm deg} {\rm \frac{erg}{g \!\cdot\! s} },~~~~
\end{eqnarray}
where $\zeta_6 = \pi^6/945$, $\alpha_{\rm EM} = 1/137$ is the electromagnetic fine-structure constant, $Y_e = Z/A $ is the number of electrons per baryon, $m_N,m_e$ are the nucleon and electron mass respectively, and
$T_8=(T/10^8\textrm{K})$ is the rescaled temperature. The function $F_{\rm deg}$ encodes the Pauli-blocking of the process due to electron degeneracy:
\beq
    F_{\rm deg} = \frac2{n_e} \int \frac{d^3 \mathbf{p}}{(2 \pi)^3} f_{e^-} (1-f_{e^-}),
\eeq
where $f_{e^-} = [e^{(E-\mu)/T}+1]^{-1}$ is the $e^-$ distribution function.
We find a good numerical approximation of $F_{\rm deg}$:
\alg{
    F_{\rm deg} &= \frac{1}{2} \left[ 1-\tanh  f(\rho,T) \right] \\ f(\rho,T) &=
    a \log_{10} \left[ \frac{\rho}{\text{g cm}^{-3}}\right] -b \log_{10}\left[ \frac{T}{\rm K}\right] +c,
}
for coefficients $a=0.973$, $b=1.596$, and $c=8.095$. At low densities and temperature, such as during stellar helium burning, the electrons are nondegenerate, and indeed we find $F_{\rm deg} \approx 1$.

The specific energy loss due to bremsstrahlung  $e+ (Z,A) \to e + (Z,A) + a$ is expected to become more important at higher densities, when $F_{\rm deg}$ falls below 1. However, this process also depends on the nucleon degeneracy. Again assuming that the electrons are nonrelativistic, the axionic bremsstrahlung rate in the non-degenerate (ND) and degenerate (D) regimes is \cite{Raffelt:1994ry}
\begin{eqnarray}
    \label{eq:Q_bremm}
    &&\mathcal Q_{b,{\rm ND}} \!=\! \frac{128}{45} \frac{\alpha_{\rm EM}^2 \alpha_{ae} \rho T^{5/2}}{\sqrt{\frac\pi2} m_N^2 m_e^{7/2}} F_{b,{\rm ND}} \! \simeq \! 0.58 \alpha_{26} {\rm \frac{erg}{g \! \cdot \! s} } \rho_3 T_8^{5/2} F_{b,{\rm ND}} \nonumber
    \\ 
    &&\mathcal Q_{b,{\rm D}} \!=\!\frac{\pi^2}{15}\frac{Z^2}{A}\frac{\alpha_{\rm EM}^2 \alpha_{ae} T^4}{m_N m_e^2} F_{b,{\rm D}} \simeq 10.8\, \alpha_{26} {\rm \frac{erg}{g \! \cdot \! s} }  T_8^4 F_{b,{\rm D}}.
\end{eqnarray}
In \Eq{eq:Q_bremm} we have defined $\rho_3 = \rho/(10^3{\rm g/cm^3})$, $F_{b,{\rm ND}}  = Z(1+Z)/A$ for metallicity $Z$,
and, to second order in the velocity at the Fermi surface $\beta_F = p_F/E_F$,
\beq \notag
F_{b,{\rm D}} = \frac{2}{3} \log\left(\frac{2+\kappa^2}{\kappa^2} \right) + \left[ \left(\kappa^2 +\frac25 \right) \log\left( \frac{2+\kappa^2}{\kappa^2} \right) - 2\right] \frac{\beta_F^2}3,
\eeq
where the Debye angle is $\kappa^2 = k_S^2/(2p_F^2)$ and the (dimensionful) Debye momentum is
\beq \label{debye-momentum}
k_S^2 = 4\pi \alpha_{\rm EM}/T\times \sum_i n_i Z_i^2,
\eeq
where the sum runs over both the electrons and the ions in the plasma.

The total specific energy loss rate from electrophilic axion emission is \cite{Raffelt:1994ry}
\beq
\mathcal Q_{ae} = \mathcal Q_{\rm sC} + ( \mathcal Q_{b,{\rm ND}}^{-1} + \mathcal Q_{b,{\rm D}}^{-1})^{-1}.
\eeq
At densities such as those encountered by massive stars in the helium burning phase, we find that the semi-Compton process dominates $\mathcal Q_{ae}$.

\subsection{Photophilic Axion}

We are similarly interested in the ``photophilic axion''. The Lagrangian involving an axion $a$ that interacts with photons is
\alg{ \label{Lagam}
    \cL &= \cL_{\rm SM} + \frac{\alpha_{\rm EM}}{8\pi} \frac af F_{\mu \nu} \widetilde F^{\mu\nu} - \frac12 m_a^2 a^2
    \\&= \cL_{\rm SM} - \frac14 g_{a\gamma} a F_{\mu \nu} \widetilde F^{\mu\nu}  - \frac12 m_a^2 a^2.
}
In \Eq{Lagam}, $g_{a\gamma}= \alpha_{\rm EM}/2\pi f$ is a dimensionful coupling which can arise from shift-symmetric derivative interactions with heavy electrically charged particles of mass scale $f \gg$ TeV. In \Eq{Lagam}, $\widetilde F^{\mu\nu}$ is the dual of the electromagnetic field strength $F_{\mu\nu}$.
For ease of presenting results, we rescale $g_{a\gamma}$ to the rough value to which massive stars are sensitive, $g_{10} \equiv g_{a\gamma}/(10^{10}\GeV^{-1})$.

The energy loss rate from a photophilic axion due to the Primakoff process is \cite{Raffelt:1990yz, Choplin:2017auq,Choplin:2017auq}
\alg{ \label{Q-ephil-ax}
\cQ_{a\gamma} &= \frac{g_{a\gamma}^2T^7}{4\pi^2 \rho} \left(\frac{k_S}{2T} \right)^2 f[\left(k_S/2T\right)^2 ]
\\  & \simeq 283.16 {\rm \frac{erg}{g \! \cdot \! s} } g_{10}^2 T_8^7 \rho_3^{-1} \left(\frac{k_S}{2T} \right)^2 f[\left(k_S/2T\right)^2 ],
}
where the function $f$ is \cite{Raffelt:1990yz}
\begin{eqnarray}
 \label{raffelt-f}
f(y^2) =\! \int\limits_0^\infty  \frac{dx}{2\pi} \left[ (x^2+y^2) \ln\left(\!1+\frac{x^2}{y^2}\right) - x^2 \right] \frac{x}{e^x-1},~~~~~
\end{eqnarray}
which we approximate as in \cite{Friedland:2012hj}. The product $y^2 f(y^2)$ goes to zero at small $y$, which in our case means that photophilic axion emission is screened at high temperatures and low densities. Using \Eq{debye-momentum}, the quantity
\begin{equation} \label{Debye-scale}
    \left(\frac{k_S}{2T}\right)^2=0.166\frac{\rho_3}{T_8^3}\sum_{j} Y_j Z_j^2,
\end{equation}
where $j$ ranges over ions and electrons \cite{Friedland:2012hj,Choplin:2017auq}.
We point out that the $T$ dependence in \Eq{Q-ephil-ax} is softened by the screening effects in \Eq{Debye-scale}, and the explicit $\rho$ dependence is cancelled by the $\rho$ dependence in \Eq{Debye-scale}, such that $\cQ_{a \gamma}$ depends on $\rho$ only from the integral function defined in \Eq{raffelt-f}.

\subsection{Hidden Photons}

Finally, we consider a hidden photon which kinetically mixes with the SM photon:
\begin{equation}
    \mathcal{L} = \cL_{\rm SM} -\frac{1}{4}F'_{\mu \nu} F'^{\mu \nu} - \frac{\epsilon}2 F'_{\mu \nu} F^{\mu \nu} - \frac{m_{A'}^2}{2} A'_\mu A'^\mu
\end{equation}
where $A'$ is the hidden photon with mass $m_{A'}$ and $\epsilon$ is the kinetic mixing parameter. 

In an electromagnetic plasma, the SM photon dispersion relation is altered; at small momentum, it appears to have a mass set by the plasma mass \cite{Braaten:1993jw}
\beq \label{plasma-mass}
\omega_p\simeq \sqrt{\frac{4\pi \alpha_{\rm EM} n_e}{m_e}} \simeq 654{\rm eV} \sqrt{\frac{Z}{A} \rho_3}.
\eeq
(In the limit of a dilute, nonrelativistic $e^+ e^-$ plasma, the plasma mass is smaller than the Debye momentum defined in \Eq{debye-momentum} by a factor $\sqrt{T/m_e}$ \cite{Raffelt:1996wa}.) The fact that arbitrarily low-energy photons cannot be produced in this environment is a ``screen'' against the copious production of hidden photons of mass $m_{A'} < \omega_p$. The dependence of the loss rate on the parameters $\ep$ and $m_{A'}$ is determined by the polarization state of the hidden photon \cite{An:2013yfc,An:2013yua}.
In a nonrelativistic plasma and assuming that $m_{A'} < \omega_p$, the specific energy loss rate is dominated by the longitudinal modes of the hidden photon, with size \cite{An:2013yfc, An:2013yua, Redondo:2013lna}
\alg{\label{eq:Q_DP}
\cQ_{A'} \!&= \!\frac{\ep^2 m_{A'}^2}{4\pi \, \rho} \frac{ \omega_p^3 }{e^{\omega_p/T} -1} \simeq \frac{\ep^2 m_{A'}^2}{4\pi} \frac{\omega_p^2T}\rho
\\&\qquad\qquad\simeq 1.8\tenx3 {\rm \frac{erg}{g \! \cdot \! s} } \frac ZA T_8 \left( \frac{\ep}{10^{-7}} \frac{m_{A'}}{\rm meV} \right)^2,
}
where in the second step we assume $\omega_p \ll T$, which is appropriate in the nonrelativistic, dilute conditions of most interest for us (although we use the full exponential expression in all conditions). During helium burning, which is of most interest for us, $Z/A=1/2$. Here we see that the explicit $\rho$ dependence is cancelled by a contribution from $\omega_p^2$, similar to the case of the photophilic axion. Thus, the rate depends on $\rho$ only in the regime in which we cannot expand the $\exp(\omega_p/T)$ in the first step; physically, this corresponds to the regime where the density is high enough that the plasma mass experiences a Boltzmann suppression. For all masses $m_{A'} \leq \omega_p$, however, the loss rate scales like $\propto \ep^2 m_{A'}^2$.

%%%%%%%%%%%%%%%%%%%%%%%%%
\section{Numerical Modelling}
%%%%%%%%%%%%%%%%%%%%%%%%%
\label{sec:MESA}

We simulate the evolution of the black hole progenitors using the stellar structure code {\tt{MESA}} version 12778 modified to include the losses due to light particle emission given in section \ref{sec:DM}. In this section we briefly describe the physical considerations behind the parameter input, and the prescriptions used to model the relevant stages of stellar evolution. We refer the reader to the {\tt{MESA}} instrument papers for more details about the code \cite{Paxton:2010ji,*Paxton:2013pj,*Paxton:2015jva} and especially \cite{Paxton:2017eie} for details of the PISN physics. Our simulation prescription for the PPISN, PISN, and core collapse follows that of references \cite{Marchant:2018kun} and \cite{Farmer:2019jed}. We use {\tt mesh\_delta\_coeff}$=0.5$ with all other parameters set to the recommended values in the \emph{test\_suite}~ {\tt ppisn}.

Our code evolves each star from the zero age helium branch (ZAHB) to either core collapse or PISN. We begin with the formation of an initial helium star of mass $M$, metallicity $Z$, and helium-4 fraction $Y({^4{\rm He}})=1-Z$ ($Y({^3{\rm He}})=0$). Following \cite{Marchant:2018kun,Farmer:2019jed} we define helium depletion as the time step at which the central helium mass fraction falls below $0.01$. The total mass of the star at this time is the mass $M_{\rm HD}$ shown in Fig.~\ref{fig:BHMG_origin}, and the mass of the carbon-oxygen core, $M_{\rm CO}$, is defined as the mass interior to the point where the helium mass fraction is larger than $0.01$ at that time. We define the mass of the black hole as the mass of bound material at core collapse, which corresponds to the mass enclosed within the radius $r$ for which the bulk material velocity at that radius is smaller than the escape velocity at that radius $v_{\rm esc}(r)=\sqrt{GM(r)/r}$.

During the helium burning phase, stars lose mass due to stellar winds. Our prescription for these losses follow that of \cite{Brott:2011ni}. In particular, $\dot{M}\propto\eta(Z/{\rm Z}_\odot)^{0.85}$ with ${\rm Z}_\odot=0.0142$. The wind efficiency parameter (clumping parameter) $\eta$ is fixed to $0.1$. Convection is modelled using mixing length theory (MLT) \cite{1968pss..book.....C} with efficiency parameter $\alpha_{\rm MLT}=2.0$: the mixing length is given by $\alpha_{\rm MLT}$ multiplied by the pressure scale height. We model semi-convection using the prescription of \cite{1985A&A...145..179L} with efficiency parameter $\alpha_{\rm SC}=1.0$. We describe convective overshooting using an exponential profile parameterized by $f_0$, which sets the point inside the convective boundary where overshooting begins, and $f_{\rm ov}$, which determines the scale height of the overshoot. We fix $f_0=0.005$ and $f_{\rm ov}=0.01$, the fiducial values used by \cite{Farmer:2019jed}. As described above, we set the nuclear burning rates to the {\tt MESA} default: these are a mixture of the {\tt NACRE} \cite{Angulo:1999zz} and {\tt REACLIB} \cite{2010ApJS..189..240C} tables.  

We verify that our grids at zero coupling agree with previous results \cite{Marchant:2018kun,Farmer:2019jed}. We find very minor differences, less than $\sim 0.5 \msun$ and well within numerical tolerances, which we attribute to using a more recent version of MESA and slightly different model resolution parameters.

%%%%%%%%%%%%%%%%%%%%%%%%%
\section{Effects of Novel Particle Losses on the Black Hole Mass Gap}
%%%%%%%%%%%%%%%%%%%%%%%%%
\label{sec:DMBHMG}

In this section we will investigate the effect of new particles by computing the black hole mass gap for grids of stellar models according to the prescription in Sec.~\ref{sec:MESA}, employing the novel particle energy loss channels identified in Sec.~\ref{sec:DM}. We will explain the physical mechanism that enables the novel energy loss channels to impact the BHMG: new losses cause faster helium burning, hastening the time to helium depletion and raising the carbon to oxygen ratio when the pair instability is encountered.

Concretely, we identify the lower edge of the mass gap by computing grids with initial masses between $20\msun$ and $90\msun$ in intervals of $1\msun$. We compute the upper edge by first computing a sparse grid with initial masses $M_{\rm in}$ between $115\msun$ and $250\msun$ with intervals of $15\msun$ and then fine-graining around the mass where black holes reappear in intervals of $1\msun$ and then $0.5\msun$ to hone in on the minimum mass above the upper edge. When presenting grids, we will plot the black hole mass as a function of the carbon-oxygen (CO) core mass rather than the initial mass because $M_{\rm CO}$ is in closer correspondence with $M_{\rm BH}$ \cite{Farmer:2019jed}. We emphasize that both the $M_{\rm in}$ and $M_{\rm CO}$ are unobservable, and are used for visualization and comparative purposes only. Our fiducial choice of metallicity is $Z=10^{-5}$. Lower metallicity stars form heavier black holes due to the reduction in wind loss, implying that these objects are responsible for the lower edge of the mass gap, which is our primary interest. We find that the upper edge of the mass gap is fairly robust to metallicity. 

Our fiducial parameter choices when investigating novel particle losses are as follows: 
\begin{itemize}
    \item {\bf Electrophilic Axion:} The coupling constants of interest for the electrophilic axion are $\alpha_{26}= 1$--$100$. A low mass axion with coupling in the upper portion of this range could potentially explain the recent XENON1T excess \cite{Croon:2020ehi, Aprile:2020tmw}. This parameter space is potentially constrained by inferred cooling rates of white dwarfs and red giants in globular clusters \cite{DiLuzio:2020jjp}, but unexplored degeneracies with parameters such as the population metallicity \cite{Gao:2020wer} or the age of the stars can reduce or remove this tension entirely. However, it is worth noting that the bound on electrophilic axions from the sun is relatively weak, constraining only $\alpha_{26} \lesssim 4000$ \cite{Redondo:2013wwa}. The constraints due to the number of relativistic degrees of freedom in the early Universe are reheat temperature-dependent and not constraining at this time, although they could provide another complementary future signal \cite{Millea:2020xxp, Arias-Aragon:2020qtn}.
    \item {\bf Photophilic Axion:} The coupling constants of interest for the photophilic axion are $g_{10}\sim \cO(1)$. A low mass axion with this coupling could potentially explain the recent XENON1T excess through inverse Primakoff absorption \cite{Aprile:2020tmw, Gao:2020wer, Dent:2020jhf}. This parameter space is probed by stellar population synthesis with similar caveats as in the electrophilic case, but is potentially also constrained by direct measurement from the CAST experiment \cite{Anastassopoulos:2017ftl}. However, the CAST experiment loses sensitivity to masses $m_a \gtrsim 0.02$eV due to a loss of magnetic field coherence. An axion mass of $m_a = 10$eV is untested by the CAST experiment, but is still produced in the core of the sun, leading to constraints from a global fit to stellar data of $g_{10} \leq 4$ \cite{Vinyoles:2015aba}. Again, constraints from the number of light degrees of freedom in the early Universe are not constraining now but could potentially provide another complementary future signal \cite{Millea:2020xxp, Arias-Aragon:2020qtn}.
\begin{figure}[t]
    \centering
   { \includegraphics[width=0.49\textwidth]{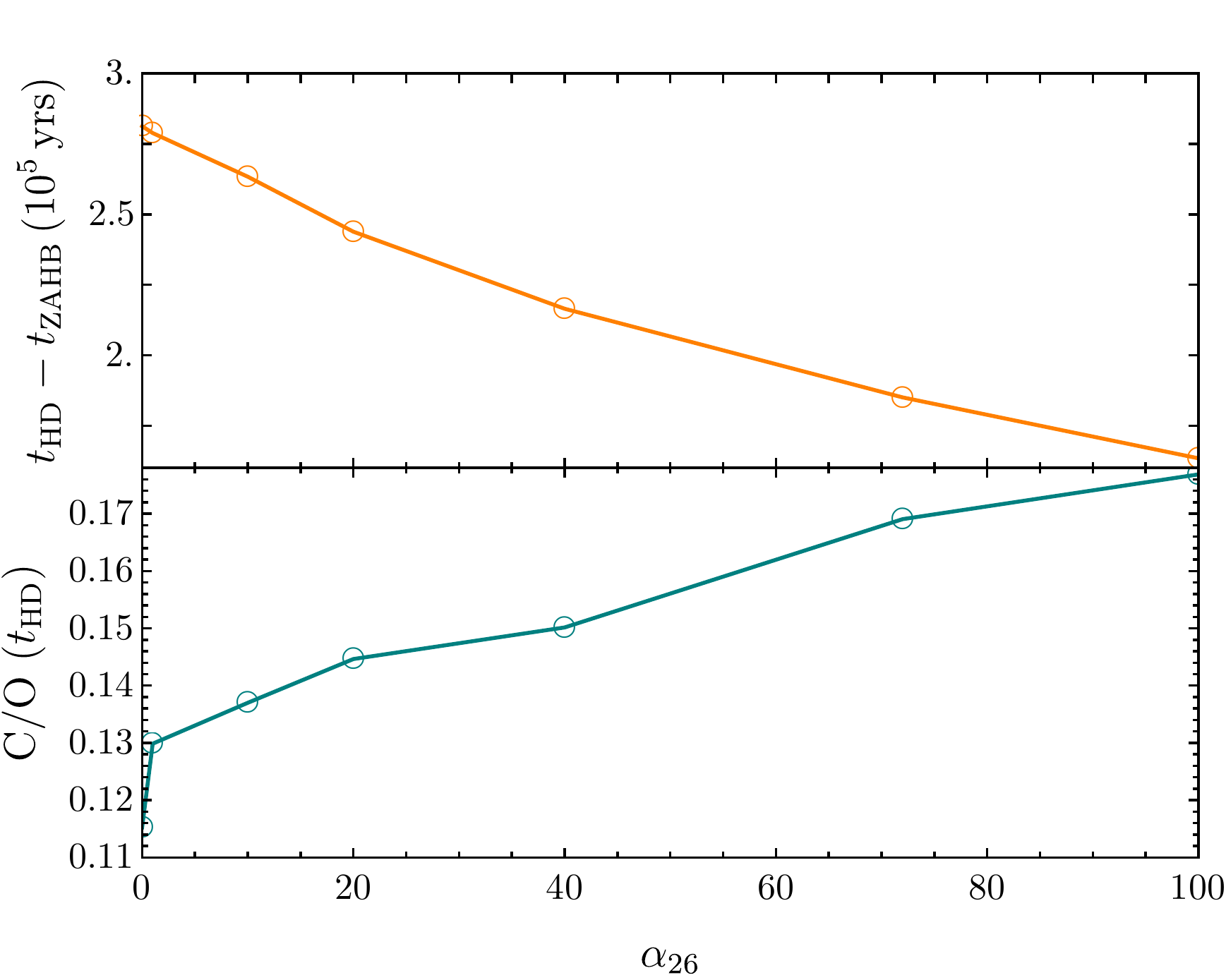}}
    \caption{
    \textbf{Top:} The time between ZAHB and helium depletion for a star of initial mass $63{\rm M}_\odot$ and $Z=10^{-5}$ as a function of $\alpha_{26}$. 
    \textbf{Bottom:} The ratio of ${}^{12}\rC$ to ${}^{16}\rO$ at helium depletion as a function of $\alpha_{26}$ for the same star.
    }
    \label{fig:HD_figs}
\end{figure}
    \item {\bf Hidden Photons:} The coupling constants of interest for hidden photons are $\ep m_{A'} \gtrsim \cO(10^{-9})$eV $\iff \ep\gtrsim 10^{-7}$ for $m_{A'}=0.01$eV. Longitudinal hidden photons are resonantly produced when the plasma frequency $\omega_p$ given in \Eq{plasma-mass} exceeds the hidden photon mass $m_{A'}$. In the cores of population-III stars, the helium cores we simulate have initial central densities of order $100$ g/cm${}^3$ and evolve towards higher densities, so we find $\omega_p \gtrsim 100 {\rm keV}$. When we consider the effects of free streaming losses in this section, we will set $m_{A'}=0.01$eV in order to have the hidden photon production active throughout the star's entire volume for its entire evolution. Of course, in the regime $\omega_p\geq m_{A'}$ the loss rate depends only on the product $m_{A'}^2\ep^2$ (see equation \eqref{eq:Q_DP}), implying that our analysis applies to rescaled values of $\ep\times(m_{A'}/$eV) over large range of parameter space meV $\lesssim m_{A'} \lesssim$ keV. Requiring that the hidden photon not overwhelm the entire luminosity of the sun leads to the requirement $\ep m_{A'} \leq 1.4\tenx{-11}$ eV \cite{An:2013yfc}; by considering effects of losses in the core on solar neutrinos, \cite{Redondo:2013lna} find the more stringent constraint $\ep m_{A'} \leq 0.4\tenx{-11}$eV. As a result, it appears that the model space probed here is not phenomenologically viable in its simplest incarnation, though our results are interesting to probe a different kind of $T$ and $\rho$ dependence for a novel loss mechanism.
\end{itemize}
We consider larger masses and different couplings when we discuss the new particle production instability in the penultimate section.

In order to understand the effects of new particle losses on the black hole mass gap, we must first understand how these losses effect individual stars. We will use the electrophilic axion to exemplify this, though the impacts on the stars will be completely analogous for photophilic axions and hidden photons. The primary effect of new particle losses is to reduce the lifetime of nuclear burning. This happens because more energy is needed to compensate for the increased losses and provide the requisite pressure necessary to maintain hydrostatic equilibrium. Since the supply of nuclear fuel is finite (at fixed stellar mass), this implies a shorter lifetime. This is exemplified in the top panel of Fig.~\ref{fig:HD_figs}, where we show the time to helium depletion for a ZAHB star of initial mass $M_{\rm in}=63{\rm M}_\odot$ as a function of $\alpha_{26}$. Evidently, the lifetime is severely reduced for increasingly strong couplings.

\begin{figure}[t]
    \centering
    \includegraphics[width=0.49\textwidth]{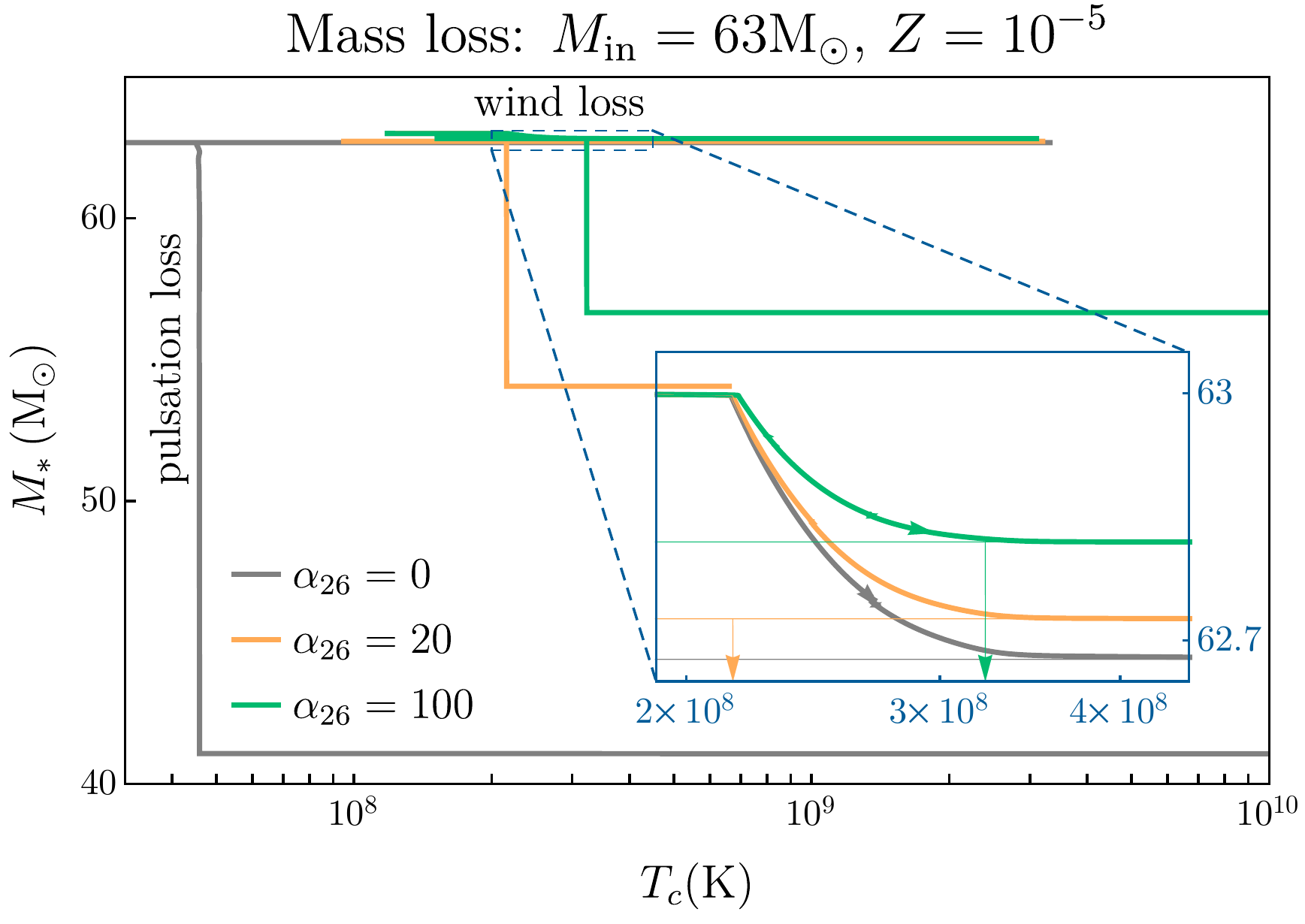}
    \caption{The mass as a function of central temperature for a star of initial mass $63{\rm M}_\odot$ for the electrophilic axion with $\alpha_{26}=0,20$  
    and 100. The inset shows the portion of the stars evolution where mass is lost to stellar winds.
    }
    \label{fig:ML}
\end{figure}

There are two important consequences of the reduced helium burning lifetime. First, the amount of mass lost due to winds is reduced since winds are active for a shorter period of time. Second, the ratio of ${}^{12}\rC$ to ${}^{16}\rO$ at helium depletion is reduced. This is because the reaction ${}^{12}\rC(\alpha,\,\gamma)^{16}\rO$, which occurs during helium burning, has less time to operate. This is exemplified in the bottom panel of Fig.~\ref{fig:HD_figs} for the same $M_{\rm in}=63{\rm M}_\odot$ star above. Oxygen is the fuel for the PPISN, and increasing the amount of carbon suppresses its burning. The consequence of the increased ${}^{12}\rC$ to ${}^{16}\rO$ ratio is then to suppress the PPISN, reducing the mass lost.

\begin{figure}[t]
    \centering
    {\includegraphics[width=.49\textwidth]{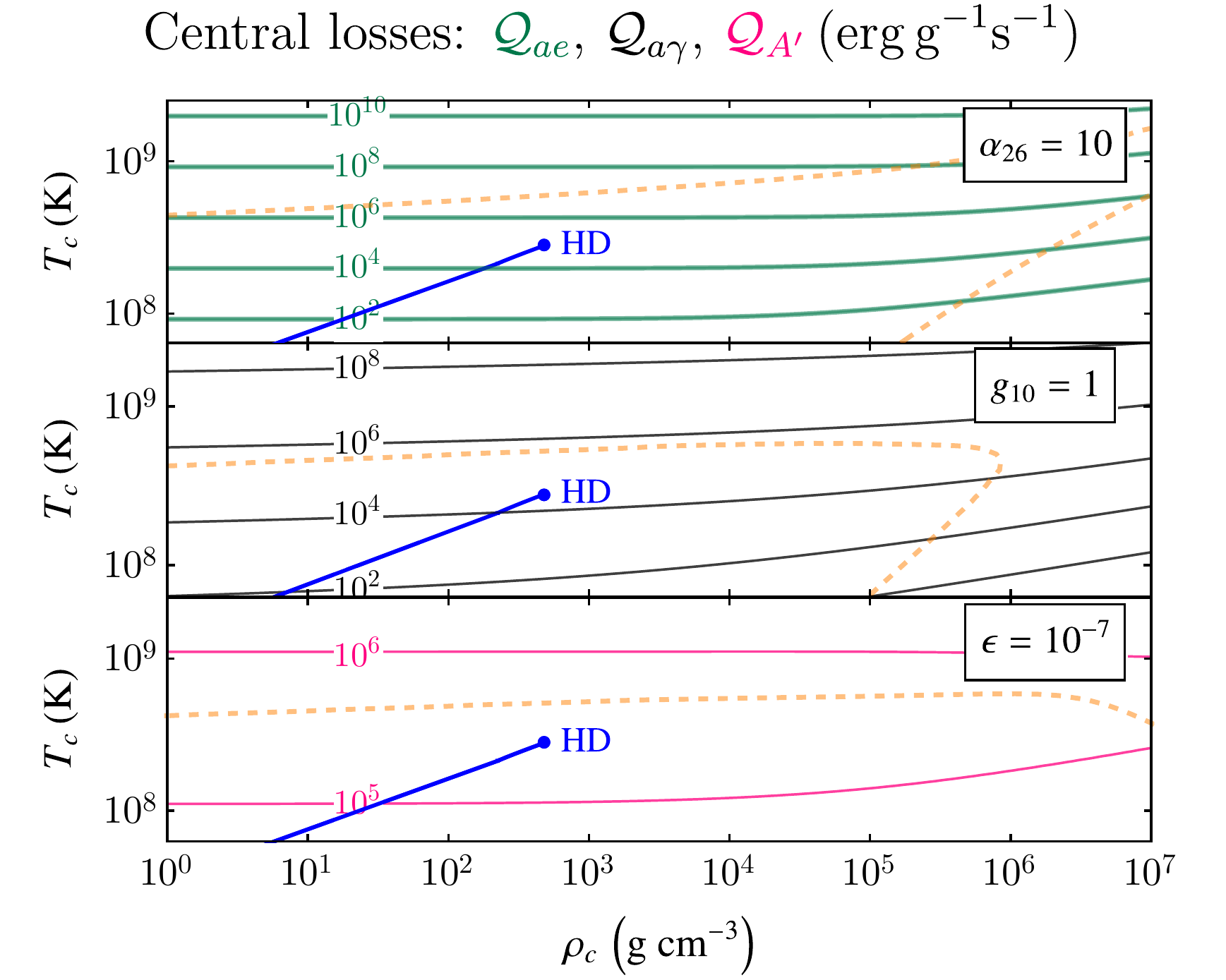}}
    \caption{Loss contours given by \eqref{eq:Q_SC}, \eqref{Q-ephil-ax}, and \eqref{eq:Q_DP} with $Z/A=1/2$ in the $T_c-\rho_c$ plane, with a sample track of a $M_{\rm in}=55 {\rm M}_\odot$, $Z=10^{-5}$ star followed until core helium depletion (HD). The dashed orange line denotes ${\mathcal Q}_i/{\mathcal Q}_\nu =1$.
    }
    \label{fig:loss_rates}
\end{figure}

These consequences are demonstrated in Fig.~\ref{fig:ML} where we plot the mass of a star with initial mass $M_{\rm in}=63{\rm M}_\odot$ as a function of its central temperature for three different values of $\alpha_{26}$. We use central temperature rather than age because, as discussed above, the age is a strong function of $\alpha_{26}$ whereas the stars all follow the same trajectory in the $T_c$--$\rho_c$ plane until the onset of the PPISN; although temperature evolves erratically as a function of time in this plot, the stellar mass monotonically decreases.
We see evidence of both the reduced mass loss from winds and the reduced explosive ability of the star in Fig.~\ref{fig:ML}. Stars with higher values of $\alpha_{26}$ lose less mass during the wind phase, as shown by the different asymptotes in the right-hand side of the inset. Yet by far a more important\footnote{This is not necessarily the case at higher metallicity. Since the wind losses scales like $\dot M\propto (Z/{\rm Z}_\odot)^{0.85}$, the shorter helium burning phase has a larger impact on higher-metallicity stars.} effect is the suppression of the PPISN. From the figure, one can see that all three stars undergo a single pulse. The star with no axion losses ($\alpha_{26}$) sheds over $20{\rm M}_\odot$ of material, whereas the stars with larger couplings only shed $\approx 7{\rm M}_\odot$. 
Similar effects are found for photophilic axions and hidden photons, but we refrain from showing these in the interest of staving repetition. In all cases, the observable consequence of the novel energy losses is that the stars collapse to form heavier black holes, an effect which increases with the magnitude of the extra dissipation.

Before presenting results, we comment on the differences between the three models we study. These models differ in their parametric dependence on the temperature and density of the star.
We show the impact of the temperature dependence by plotting contours of constant loss rate in Fig.~\ref{fig:loss_rates}. The blue track shows the evolution of a star of initial mass $M_{\rm in}=55{\rm M}_\odot$ ($Z=10^{-5}$ with no new particle emission) from the ZAHB to the onset of Carbon ignition. We show as a dashed orange line the contour along which neutrino losses $\cQ_\nu$ (with $\cQ_\nu$ from \cite{Itoh1996}) become dominant over the novel particle emission; neutrino losses dominate at higher temperatures. The appearance of this contour slightly beyond the point of helium depletion validates our heuristic argument for assessing the impact of novel losses based on shortening the time to helium depletion: if neutrino losses dominated well before this time, then adding a new emission mechanism would not impact the stellar evolution. Thus, the novel particles we discuss offer a qualitatively new energy loss channel for the star in a phase of its life where it would otherwise not experience such losses.
This figure also shows the difference in the temperature scaling of the novel particle losses. At low densities, the loss rates given in \eqref{eq:Q_SC}, \eqref{Q-ephil-ax}, and \eqref{eq:Q_DP} scale like $\cQ_{ae} \propto T^6$, $\cQ_{a\gamma} \propto T^4$, and $\cQ_{A'} \propto T^1$ (each rate depends on density through screening, but this is unimportant for the present comparison).  The top panel shows the effect of an electrophilic axion with $\alpha_{26} =10$, the middle panel shows the effect of a photophilic axion with $g_{10}=1$, and the bottom panel shows the effect of a hidden photon with $\ep=10^{-7}$. The stellar track we show ends with $\cQ \simeq 10^5$erg/g/s in all cases, but in the electrophilic axion case this reflects an increase of over five orders of magnitude in rate from the moment of  initialization, while in the hidden photon case the rate changes by only an order of magnitude. 
The significance of this difference in scaling is that if we increase the loss rates by increasing the couplings of the new particles, we find that we increase the {\it time-integrated loss} due to the hidden photon by more than the corresponding increase in the axion cases.
These insights are reflected in the scaling of the edges of the BHMG with the different coupling constants, to which we now turn.

\begin{figure}[t]
    \centering
    {\includegraphics[width=.49\textwidth]{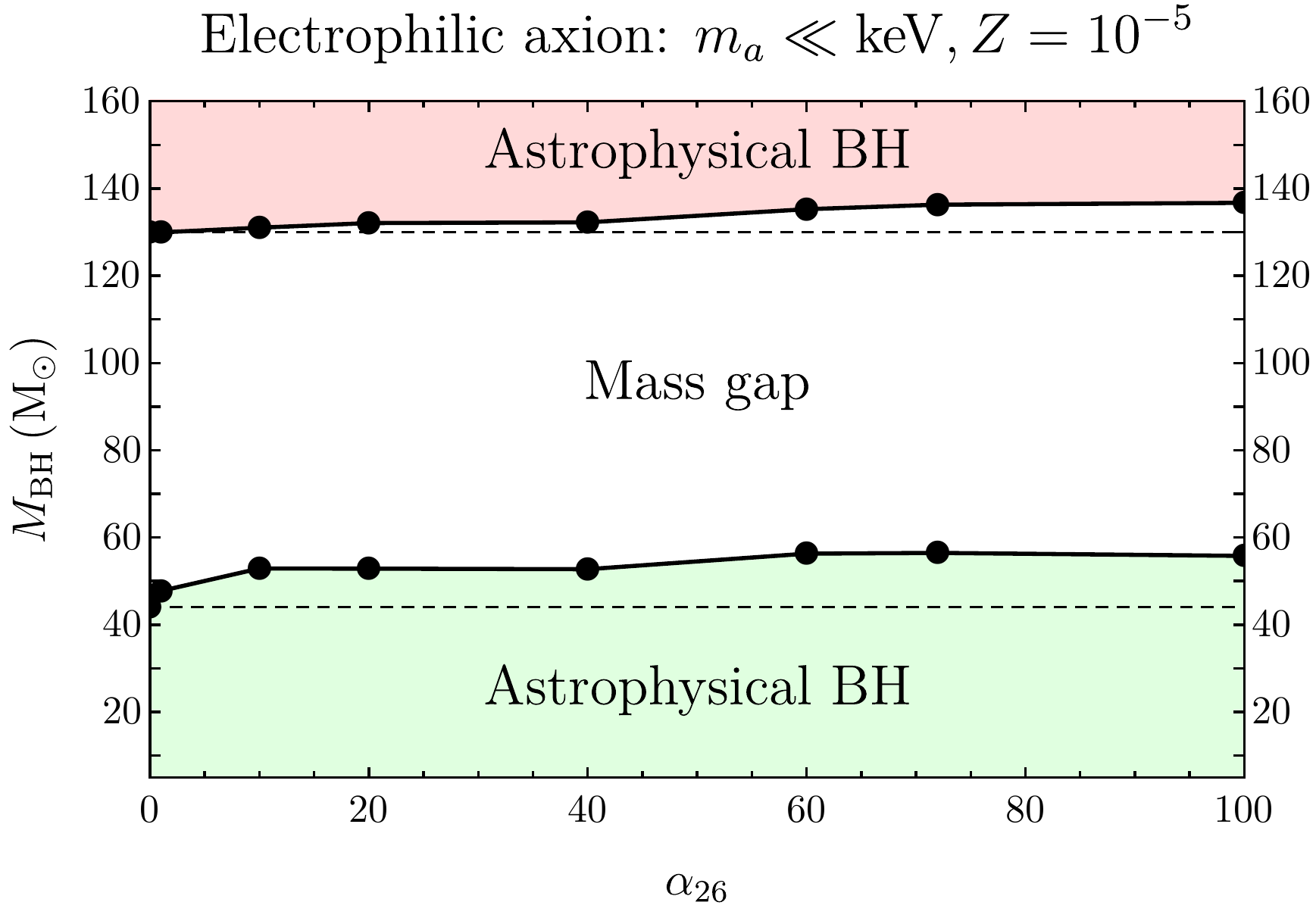}}
    \caption{The black hole mass gap predicted when electrophilic axion losses are included in stellar evolution. The plot shows the maximum black hole mass below the mass gap and the minimum mass above the mass gap as a function of $\alpha_{26}$. The stars had initial metallicity $Z=10^{-5}$. The gray dashed lines indicate the fiducial black hole mass gap predicted by the Standard Model.}
    \label{fig:grid_AE}
\end{figure}

\begin{figure*}[t]
    \centering
    {\includegraphics[width=1.\textwidth]{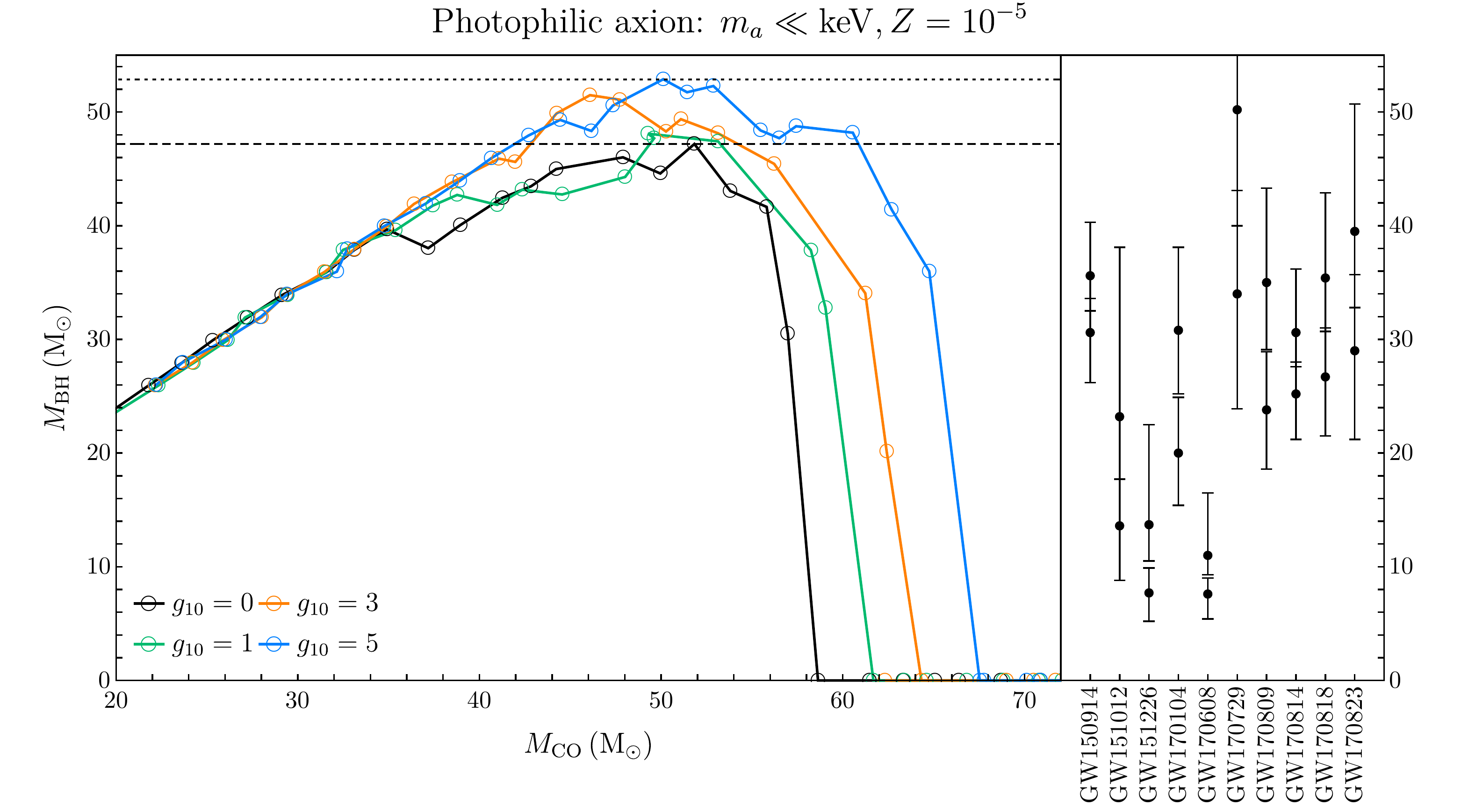}}
    \caption{Final black hole mass as a function of CO core mass when losses due to photophilic-axions are included. We explore various values of $g_{10}$ indicated in the figure. The dashed line corresponds to the lower edge of the mass gap for  $\alpha_{26}=0$, consistent with \cite{Farmer:2019jed}; the dotted line is the maximum black hole mass below the BHMG for the couplings we simulate in this model. The initial black hole masses inferred from the first 10 binary black hole mergers observed in the first two LIGO/Virgo observing runs are shown in the right hand panel.}
    \label{fig:grid_axions}
\end{figure*}
\begin{figure}[h]
    {\includegraphics[width=.49\textwidth]{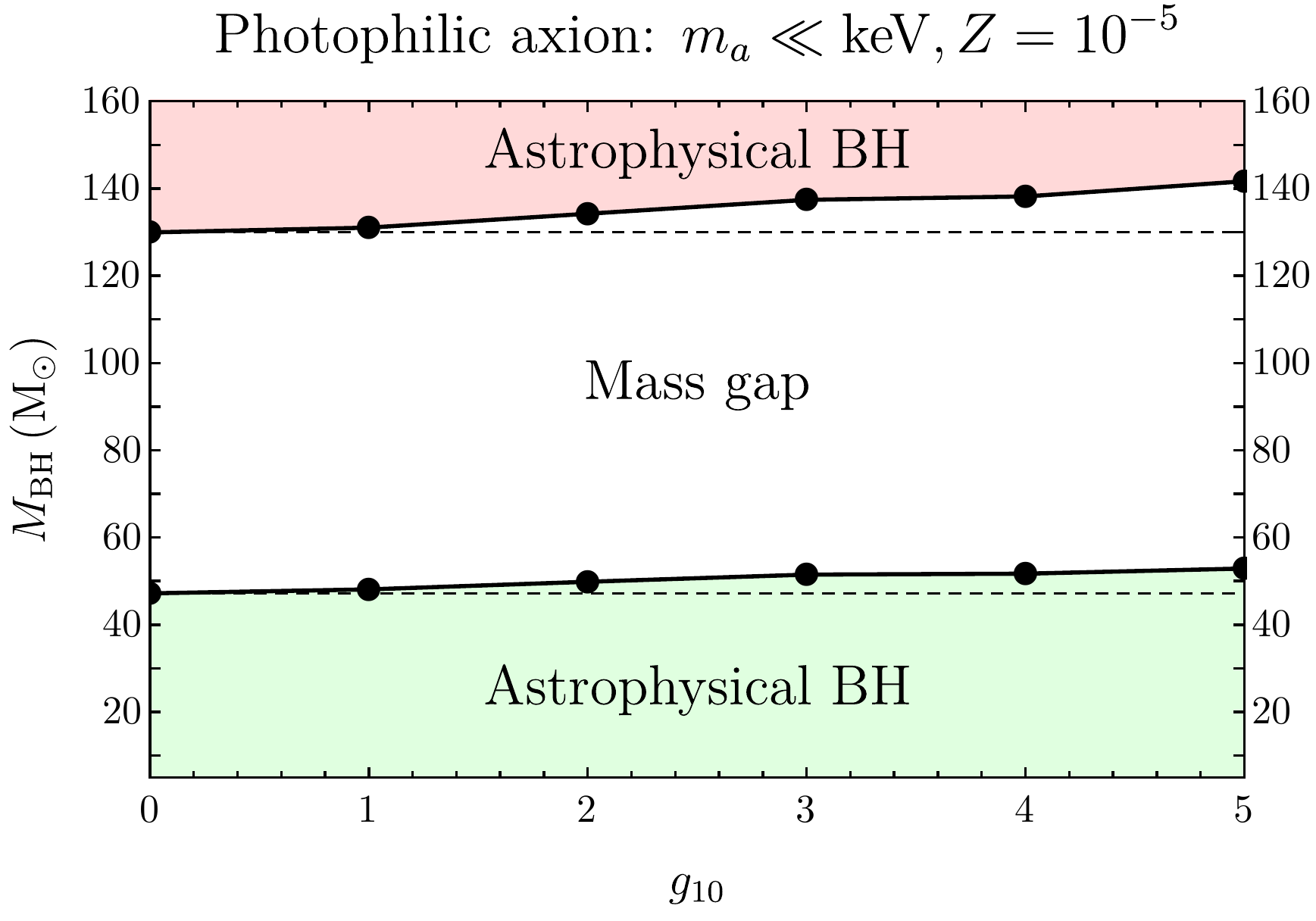}}
    \caption{The black hole mass gap with a photophilic axion as a function of $g_{10}$. Black circles correspond to individual stellar models with metallicity $Z=10^{-5}$. The gray dashed lines indicate the black hole mass gap with Standard Model particles only. }
    \label{fig:massgap_PA}
\end{figure}

\begin{figure*}[t]
    \centering
    {\includegraphics[width=1.\textwidth]{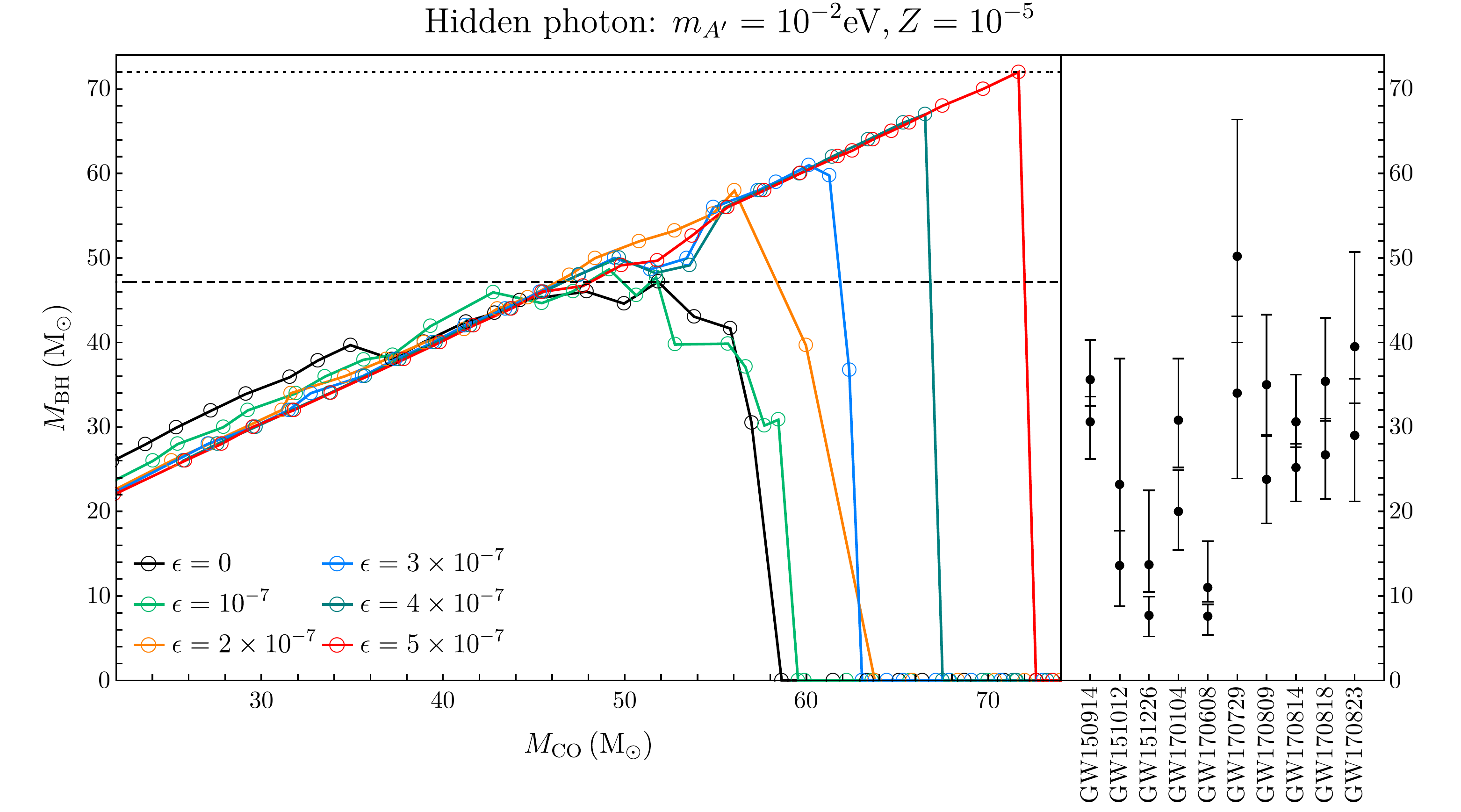}}
    \caption{Final black hole mass as a function of the carbon-oxygen core mass of its progenitor when losses due to hidden photons of mass $m_{A'}=0.01$eV and kinetic mixing with the photon $\ep$ indicated in the figure are included. As before, the black dashed line shows the lower edge of the mass gap without new particle losses, and the dotted line is the maximum black hole mass below the BHMG for the couplings we simulate in this model. The initial black hole masses inferred from the first 10 binary black hole mergers observed in the first two LIGO/Virgo observing runs are shown in the right hand panel.
    }
    \label{fig:grid_DP_z_div_solar}
\end{figure*}
\begin{figure}[h]
    {\includegraphics[width=.49\textwidth]{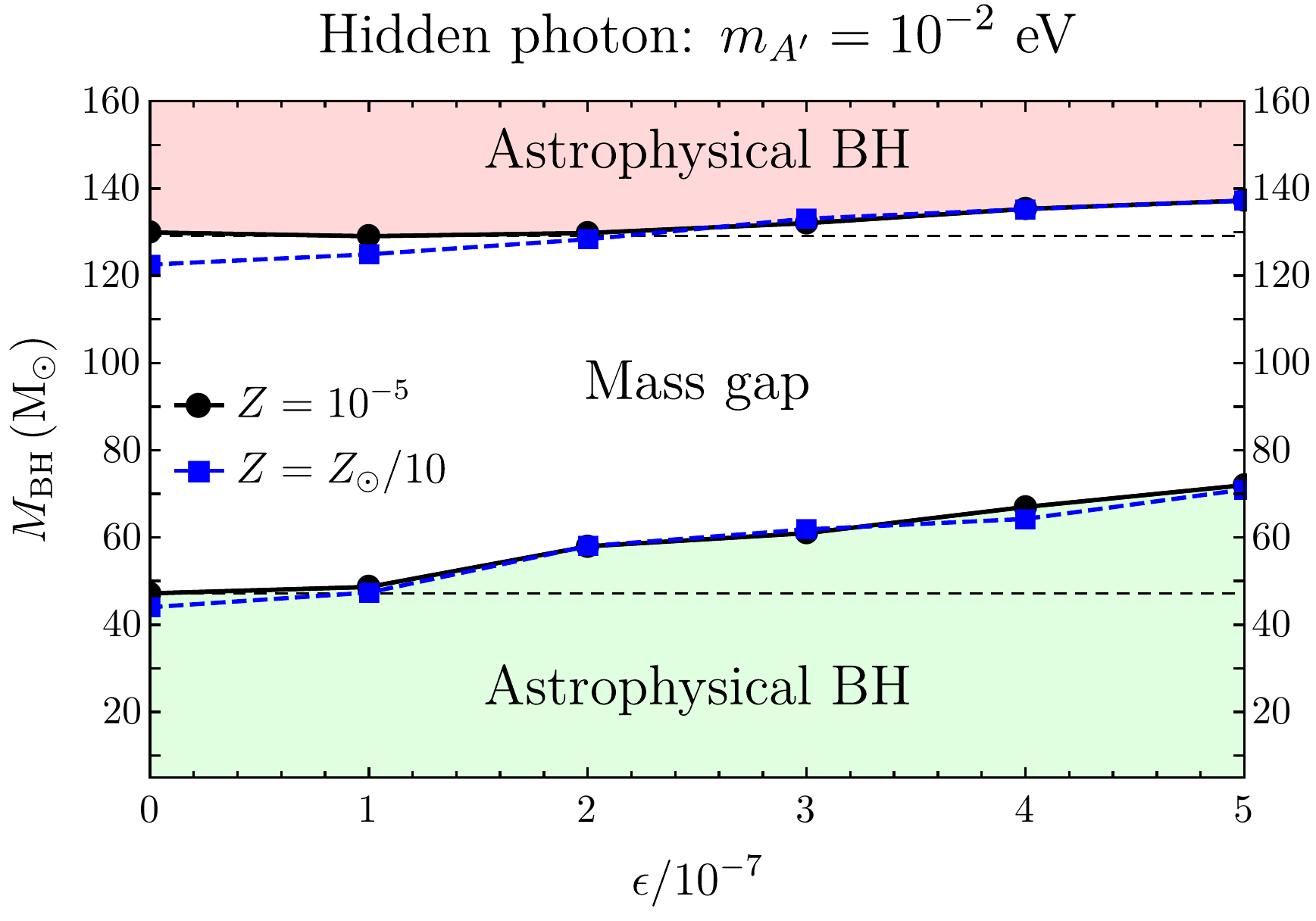}}
    \caption{The black hole mass gap with a hidden photon of mass $m_{A'}=0.01$eV as a function of $\ep$. Black circles correspond to $Z=10^{-5}$ models and blue squares to ${\rm Z}_\odot/10$. The gray dashed lines indicate the fiducial black hole mass gap predicted by the Standard Model. }
    \label{fig:massgap_DP}
\end{figure}

For the photophilic axion and the hidden photon, will describe the consequences of novel particle production on the BHMG by plotting $M_{\rm BH}$ as a function of $M_{\rm CO}$, following \cite{Farmer:2019jed}, for a grid of initial masses as described above and a variety of different coupling choices. For all three models we will show summary results, taking the heaviest black hole on the low end of the gap and the lightest black hole on the high end of the gap from each of our grids of stellar masses and plotting these as a function of the coupling parameter to new physics.

First we describe the impact of new losses on the black hole population in the case of an electrophilic axion. This model was originally studied in \cite{Croon:2020ehi}; here, we repeat the conclusions of that study and add a discussion of heavier progenitors.
We show the (two-sided) BHMG as a function of the coupling $\alpha_{26}$ in Fig.~\ref{fig:grid_AE}.
Evidently, stronger couplings lead to heavier black holes, and the location of the lower edge of the mass gap moves from $47{\rm M}_\odot$ for $\alpha_{26}=0$ (\ie,~the Standard Model prediction) to $56{\rm M}_\odot$ for $\alpha_{26}=100$. The same effect can be seen for the upper edge of the mass gap, corresponding to black holes that do not experience a PISN because the pair instability is quenched by the photodisintegration of heavy elements. Consequently this upper edge moves to higher masses by $\sim8{\rm M}_\odot$. Despite changes in the values of the edges of the BHMG, the {\it size} of the mass gap remains similar for all values of the new physics coupling, as anticipated by \cite{Farmer:2020xne, Ezquiaga:2020tns}. These shift in the edges of the BHMG exceed the uncertainties and inherent scatter of the processes discussed above, meaning that novel particle losses have the potential to produce black holes in new mass ranges, thereby opening the exciting possibility that LIGO/Virgo could be detectors of new particle physics via black hole population studies.

Turning to photophilic axions, the black hole mass distribution for low mass progenitors (the lower edge) is shown in Fig.~\ref{fig:grid_axions}. As expected, we find similar effects, namely that increasing the rate of Primakoff losses by raising $g_{10}$ leads to higher mass final black holes \ie,~the mass gap once again shifts to higher masses. In this case we do not see such large final black hole masses as we did for electrophilic axions. This is because, motivated by Solar constraints \cite{Vinyoles:2015aba}, we did not explore very large values of the coupling $g_{10}$: as compared to the electrophilic axion, in which we explored a range of 100 in the rate, we explore only a relative factor of 25 in rate for the photophilic axion. We also show in Fig.~\ref{fig:massgap_PA} the complete mass gap including the upper edge for the case of the photophilic axion. As in the case of the electrophilic axion, the width of the BHMG does not change substantially when including the photophilic axion, although we do find some possible evidence that the BHMG widens by $\sim 5\msun$ at larger $g_{10}$.
Despite the relatively compressed range of values of $g_{10}$ that we simulate, we do find that a photophilic axion with a coupling $g_{10}=5$, in modest tension with solar data \cite{Vinyoles:2015aba}, will change the BHMG at a level exceeding the known uncertainties and scatter.
We also take this opportunity to remark that the effects of photophilic axions on population-III stars have been studied by \cite{Choplin:2017auq} (without the inclusion of pulsations) using the {\tt{Geneva}} code \cite{eggenberger2008geneva}. Interestingly, \cite{Choplin:2017auq} find a novel feature for $g_{10}\ge1$, where trajectories in the $T_c$--$\rho_c$ plane curve to higher densities at fixed $T_c$, indicating increased core contraction. This raises the possibility that photophilic axions may cause some stars to avoid the instability region completely, which would in its most spectacular manifestation lead to the absence of a mass gap. However, despite an extensive search---varying the various input parameters and updating our numerical prescriptions to match those of \cite{Choplin:2017auq}---we have been unable to replicate this behavior for any of the three models studied here.

Finally, we study the effects of hidden photons on the BHMG. The final black hole population formed below the PISN is shown in Fig.~\ref{fig:grid_DP_z_div_solar} and the complete mass gap including the upper edge in Fig.~\ref{fig:massgap_DP}. In this case, the effects of the losses are more pronounced and show a much steeper and tighter correlation with increasing coupling than in the previous cases considered. This derives from the temperature scaling noted above: because $\cQ_{A'} \propto T_c$, increasing the kinetic mixing parameter $\ep$ corresponds to a much larger time-integrated energy loss due to the hidden photon than the corresponding increase in $g_{10}$ does in the photophilic axion case.
Intriguingly, the PPISN is quenched completely for $\ep=5\tenx{-7}$, so the final black hole mass for those stars is approximately equal to the star's mass at helium depletion \ie,~the only effects of mass loss are due to stellar winds and, at higher masses, the PISN. Black holes as heavy as $72{\rm M}_\odot$ can be formed. These would lie very deep within the mass gap predicted by the Standard Model. Interestingly, reference \cite{Liu:2019lfc} have reported the discovery of a $\sim70{\rm M}_\odot$ black hole in the Milky Way. The Standard Model lacks a formation model for such an object, but this is not the case if hidden photons exist. It is very likely that in reality this object has a mass in the range $5{\rm M}_\odot\le M\le 20{\rm M}_\odot$ \cite{Eldridge:2019ktq,El-Badry:2019duz}, but if a similar object were to be observed in the future, new particle losses, especially hidden photons, would represent one possible formation channel. We wish to reiterate, as above, that these kinetic mixing parameters are excluded by Solar observations \cite{An:2013yfc, Redondo:2013lna}, but we think that this motivates the construction of viable models of new physics with loss rates $\cQ_{\rm new} \propto T^n$ for $n\leq 1$.

In the case of the hidden photon, we have also calculated the BHMG for $Z={\rm Z}_\odot/10$. One can indeed see from Fig.~\ref{fig:massgap_DP} that the precise location is relativity robust to changing $Z$. Since larger metallicities imply more wind loss, the upper values of the lower edge of the BHMG correspond to $Z=10^{-5}$ and the lower values of the upper edge to $Z={\rm Z}_\odot/10$.

%%%%%%%%%%%%%%%%%%%%%%%%%
\section{Massive Novel Particle Production and New Instabilities}
%%%%%%%%%%%%%%%%%%%%%%%%%
\label{sec:new_instab}

A new particle that is sufficiently massive will not escape from the star's gravitational potential well but rather will remain inside the core. In this case, the particle will not act as a loss source but will instead contribute to the EOS, provided the coupling to the Standard Model is strong enough to attain and maintain thermal equilibrium. This then allows for the possibility that the production of new particles in the cores of massive stars could give rise to a new instability. 

In order to investigate this, we have calculated the equation of state for a gas of ions, radiation, electrons (and positrons), and novel particles $X$ in thermal equilibrium. We need not restrict ourselves to a particular model, so we consider $X$ that has a mass $m_X$ and degeneracy $g_X$ ($g_X=1$ for axions, $g_X=2$ for a single Weyl fermion, $g_X=3$ for hidden photons, $g_X=4$ for a Dirac fermion). Model dependence enters via the coupling to matter, which dictates the timescale needed to attain thermal equilibrium. We will explore the ranges of couplings for well-studied models at the end of this section. The calculations of the EOS are given in Appendix \ref{sec:instab_region}. Note that we assume that the particle has no initial abundance in the star, and therefore has zero chemical potential. Models for which this is not the case, or particles charged under the Standard Model gauge group, may need a separate treatment.
\begin{figure}
    \centering
    {\includegraphics[width=.49\textwidth]{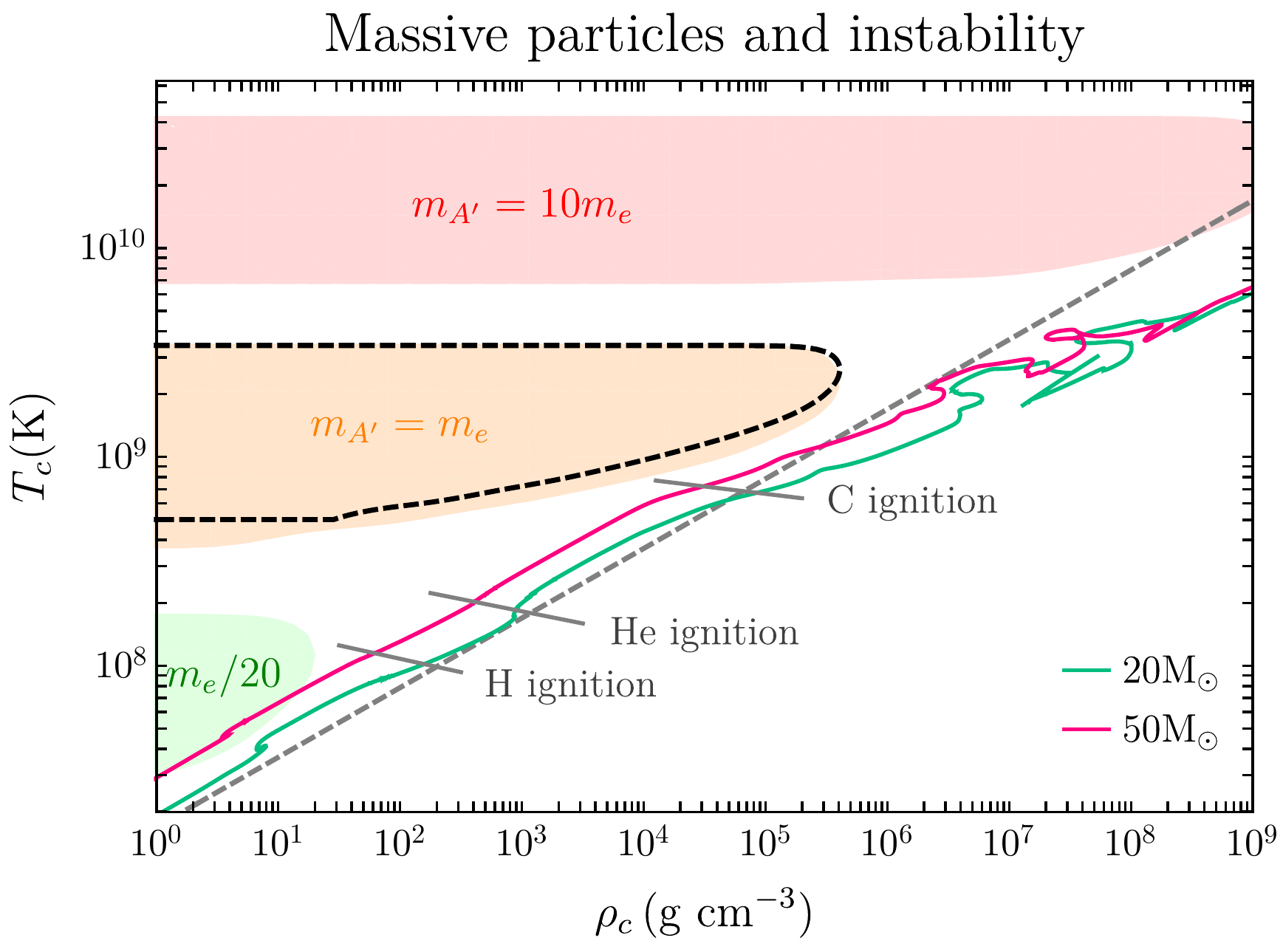}}
    \caption{The region in the $T_c$--$\rho_c$ plane where $\Gamma_1<4/3$ due to the production of new particles $X$ of mass $m_X$ indicated in the figure signifying a potential new instability. The particles have degeneracy $g_X=3$. The tracks corresponds to stars with metallicity $Z={\rm Z}_\odot/10=0.00142$ and zero age main-sequence masses indicated in the figure. The gray dashed line indicates where the radiation pressure is equal to the gas pressure, the gas pressure dominating at higher densities. The black dashed line indicates the region where $e^+e^-$ pair-instability is active.  }
    \label{fig:new_instab}
\end{figure}

The region in the $\rho_c$--$T_c$ plane where stars are unstable to $X$ production is shown in Fig.~\ref{fig:new_instab} for $g_X=3$ and various values of $m_X$. Also shown are the tracks for various massive stars beginning at the zero age main-sequence (ZAMS) indicated in the figure\footnote{These stars lose mass to winds, so the ZAMS mass is not necessarily equivalent to ZAHB mass used in preceding sections.}. One can see that the tracks do not encounter the instability region for $m_X>m_e$, but for $m_X<m_e$ it is possible for stars as light as $40\msun$ to pass through an unstable region. In particular, for $m_X\sim1$--$50$keV, hydrogen and helium burning stars may pass through the new instability.

The presence of an unstable region alone is not sufficient to ensure that the star is indeed destabilized; the new particle must couple to the Standard Model strongly enough to attain thermal equilibrium. As an example, for hidden photons the rate of the process $\gamma e^-\rightarrow A'e^-$ is proportional to $\ep^2$. Assuming the timescale for equilibration is set by
\beq
t_{A'} \simeq \Gamma^{-1}_{A'} \simeq \pL\ep^2 \sigma_T n_e e^{-m_{A'}/T_c} \pR^{-1},
\eeq
where $\sigma_T = 8\pi \alpha_{\rm EM}^2/3m_e^2$, and putting in characteristic values $m_{A'} = 100$keV, $T_c = 2\tenx8$K, and $n_e = \rho_c/m_N$ with $\rho_c = 2\tenx2$g/cm${}^3$ and $Z/A=1/2$ (characteristic of a star at the beginning of helium burning), we find that the $A'$ population equilibrates on a timescale $\sim 10^5$ years for a mixing parameter $\ep \simeq 3\tenx{-12}$. A hidden photon with this mass and mixing is much too heavy to be produced in the sun, and has a small impact on horizontal branch stars \cite{Redondo:2013lna}. Thus, it is possible that this part of parameter space is uniquely probed by the evolution of massive stars. Similarly, if we approximate the electrophilic axion equilibration timescale as
\beq
t_{ae} \simeq \pL  \frac{\alpha_{26}}{\alpha_{\rm EM}} \frac{\sigma_T}2 n_e e^{-m_{A'}/T_c} \pR^{-1},
\eeq
and taking $m_a = 100$keV with the same parameters as above, we find that the $a$ population is equilibrated after $\sim 10^5$ years for $\alpha_{26} \sim 10$. An axion of this mass will, again, have only a small impact on horizontal branch stars, so this provides a new possible signal of an electrophilic axion. Finally, the photophilic axion will equilibrate over a timescale $t_{a\gamma} \simeq (\alpha_{\rm EM} g_{a\gamma}^2 n_e)^{-1}$, and we find $t_{a\gamma} \simeq 10^5$ years for $g_{10} \simeq 10$. This, too, is likely unprobed by conventional stellar constraints.

Regrettably, the precise nature of this instability cannot be determined analytically. One possibility is a thermonuclear explosion as is the case with the pair-instability. Another possibility is that some stable cycle is reached and the ultimate result is a new type of pulsating or variable star. In the case of the pair-instability, the result is the thermonuclear burning of ${}^{16}\rO$. In our case, the instability would likely appear during hydrogen burning phase, or during the phase between hydrogen exhaustion and helium burning. Performing a detailed numerical analysis of the instability is well beyond the scope of our current work, but this analysis provides the hope that massive stars can act as a unique laboratory of sub-MeV hidden sector particles.

%%%%%%%%%%%%%%%%%%%%%%%%%
\section{Discussion and Outlook}
%%%%%%%%%%%%%%%%%%%%%%%%%
\label{sec:concs}

In this paper, we demonstrate that the black hole mass gap has the potential to become a powerful new tool for testing fundamental particle physics. New light particles that couple to luminous matter are a ubiquitous prediction in theories of physics beyond the Standard Model. Such particles can be produced in stellar interiors and act as an additional source of energy loss. 
This work shows that these losses can drastically alter the late stages of the evolution of population-III stars in two important ways. First, the lifetime of helium burning is significantly reduced, resulting in a diminished amount of mass loss due to stellar winds. Second, the pulsations that arise as a result of the pair-instability are weakened and can be quenched entirely for strong-enough couplings. This is due to an increase in the ratio of ${}^{12}\rC$ to ${}^{16}\rO$ at the end of helium burning. The ${}^{12}$C$(\alpha,\gamma)^{16}$O reaction that is active during helium burning has less time to operate, leading to the increase in the ratio of carbon to oxygen. Oxygen is the fuel for the explosive part of the pulsations, whereas carbon acts to quench it, so increasing the C/O ratio has the overall effect of suppressing the pulsations \cite{Farmer:2020xne}. The end result is that the location of the mass gap (both the upper and lower edge) is raised to higher masses.

We study this effect this for two species of new light particles -- hidden photons and axions. In the latter case, there are two potentially interesting couplings, the coupling to photons, which results in axion production via the Primakoff process, and the coupling to electrons, which results in axion emission via semi-Compton and bremsstrahlung processes. In the former case, the primary driver of light hidden photon emission is resonant production due to kinetic mixing with the Standard Model photon, although we touched on a possible impact of semi-Compton production for a sufficiently massive hidden photon. 

We observe visible departures from the Standard Model predictions when $\ep m_{A'}\gtrsim10^{-9}$eV (hidden photons), $g_{a\gamma}>10^{-10}$ GeV${}^{-1}$ (photophilic axion), and $g_{ae}\gtrsim3\tenx{-13}$ (electrophilic axion). For hidden photons, these parameters are already excluded by other stellar probes \cite{An:2013yfc, Redondo:2013lna}. For axions, the values of $g_{a\gamma}$ of interest are commensurate with the bounds from our understanding of the Sun \cite{Vinyoles:2015aba}, and the values of $g_{ae}$ are near current experimental limits \cite{DiLuzio:2020jjp, Gao:2020wer}.

For the lower edge of the black hole mass gap, the experimental constraints obtained on these couplings or the prospects for new discoveries, is strongly time-dependent. Performing a Markov Chain Monte Carlo analysis of early LIGO/Virgo data, \cite{Fishbach:2017zga} found evidence that the lower edge of the black hole mass gap lies at $~40\msun$. We postpone a similar analysis -- including the effects described in this paper -- until the release of the data from the O3 observing run, as we may anticipate up to 50 more binary mergers involving a black hole (in addition to the 10 reported presently) with improved error bars. Unlike other stellar probes, as LIGO/Virgo is upgraded to its full sensitivity and additional detectors come online, the data set will improve significantly in quantity and quality in the near future. As a result, the BHMG will come into sharp focus, and the mechanism described here will become a sensitive probe of new physics.

We also study the upper edge of the black hole mass gap. Very massive stars ($>120 \msun$ for $Z=10^{-5}$) do not experience a PISN since some of the energy from the contraction goes towards the photo-disintegration of heavy elements, which quenches the instability. As a result, the black hole mass gap ends and stars with masses $M>120\msun$ are expected to exist. We find that light particle emission raises the upper edge of the mass gap to higher masses. Reference \cite{Ezquiaga:2020tns} has recently argued that LIGO/Virgo may be sensitive to black holes with masses just above the mass gap once they are upgraded to `A+' sensitivity. Our results imply that such black holes may be heavier still, and it would be interesting to investigate if the lack of such observations, or detailed population studies, could be used to place new bounds.

Additionally, we examine the potential effects of heavy novel particles on the evolution of population-III stars. Such novel particles, if coupled strongly enough to the Standard Model, can accumulate in the cores of stars and remain in thermal equilibrium with luminous matter. In this scenario it is possible to trigger a new instability. We derive the region in temperature and density where this would apply, by a direct calculation of the equation of state for a gas of novel particles in equilibrium with radiation, ions, and electron-positron pairs. We find that stars as light as $M\sim40\msun$ could potentially encounter this instability during their hydrogen or helium burning phases. Without detailed numerical modelling we are unable to determine the exact nature of the instability, but it could potentially lead to pulsations or to stellar disruption, depending on how the core helium reacts to the star's contraction. The results of such a detailed numerical modeling may reveal a new window to new physics from massive stars.

In summary, the preliminary exploration undertaken in this work demonstrates that the black hole mass gap has the potential to become a powerful tool in the quest to find physics beyond the Standard Model. Looking ahead, as the third LIGO/Virgo observing run is concluded, the apparatus is upgraded to even higher sensitivities, and future detectors such as LIGO-India and KAGRA are coming on line \cite{Akutsu:2020his}, we expect hundreds to thousands of events per year. Identifying potential observables, such as the shift in the location of the black hole mass gap to higher masses that we have predicted here is of paramount importance for unleashing the full potential of this data, and for maximizing its discovery potential. The methods we present here can be adapted to make predictions for a variety of new physics models to capitalize on that potential.

%%%%%%%%%%%%%%%%%%%%%%%%%
\section*{Acknowledgments}
%%%%%%%%%%%%%%%%%%%%%%%%%

Many thanks to Eric Baxter, James deBoer, Jose Maria Ezquiaga, Dan Fabrycky, Jason Kumar, Kristina Launey, Jess McIver, Noemi Rocco, Istvan Szapudi, Xerxes Tata, and Adrian Thompson. We are grateful to Robert Farmer and the {\tt MESA} community for their help in answering our numerous questions. SDM would like to acknowledge the hospitality of JS at the University of Pennsylvania, where parts of this work were initiated. T\acro{RIUMF} receives federal funding via a contribution agreement with the National Research Council Canada.
Fermilab is operated by Fermi Research Alliance, LLC under Contract No. De-AC02-07CH11359 with the United States Department of Energy.

\appendix

\section{Derivation of the instability regions}
\label{sec:instab_region}
In this Appendix we will re-derive the instability region caused by electron-positron pairs following \cite{1967ApJ...148..803R}. This is found by deriving the equation of state and looking for regions where $\Gamma_1<4/3$. Next, we will extend this calculation to include possible heavy novel particles to derive the new instability region. The ultimate goal is to calculate the adiabatic index given by
\begin{equation} \label{eq:EOS1}
\Gamma_1=\frac{\rho}{P}\left(\frac{\partial P}{\partial\rho}\right)_s=\left(\frac{\partial P}{\partial\rho}\right)_T+\left(\frac{\partial P}{\partial T}\right)_\rho\left(\frac{\partial T}{\partial \rho}\right)_s
\end{equation}
with
\begin{equation}\label{eq:EOS2}
\left(\frac{\partial T}{\partial \rho}\right)_s=-\frac{\left({\partial s}/{\partial \rho}\right)_T}{\left({\partial s}/{\partial T}\right)_\rho}.
\end{equation}
We take $s$ to be the entropy per unit mass throughout.

\subsection{Ions}

The relevant thermodynamic relations for these objects are the ideal gas law and the Sackur-Tetrode equation:
\alg{
P_{\rm ions}&=\left\langle\frac{1}{A}\right\rangle\frac{n k_B T}{m_H},
\\ \quad\textrm{and}\quad s&=\left\langle\frac{1}{A}\right\rangle\frac{k_B}{m_H}\left[\frac{5}{2}+\ln\left(\frac{T^\frac32}{\rho}\right)\right]
}
where $\langle1/A\rangle$ is the average reciprocal atomic unit. One can straightforwardly take partial derivatives of these to compute the quantities in equations \eqref{eq:EOS1} and \eqref{eq:EOS2}. Note that we treat the ions as a (fully ionized) monatomic gas since we will treat the electrons separately.

\subsection{Radiation}

In this case, the relevant quantities are the radiation pressure and the radiation entropy:
\begin{equation}
P_{\rm rad}=\frac{a}{3}T^4,\quad\textrm{and}\quad s=\frac43 a\frac{T^3}{\rho}.
\end{equation}
Again, it is straight forward to compute the partial derivatives needed to compute equations \eqref{eq:EOS1} and \eqref{eq:EOS2}.

\subsection{Electrons and positrons}

We calculate the contribution to the equation of state by integrating over the Fermi-Dirac distribution and imposing charge neutrality. First, we define the quantities
\begin{equation}
C_e=\frac{1}{\pi^2}\left(\frac{m_e c}{\hbar}\right)^3,\quad \beta_e(T)=\frac{m_e c^2}{k_B T},\quad\textrm{and}\quad \phi=\frac{\mu}{k_B T},
\end{equation}
where $\mu$ is the chemical potential. From this, we can derive the pressure, density, charge density, and entropy as follows:
\begin{align}
P_e(\phi,\beta_e)&=m_e c^2C_e F_1(\beta_e,\phi)\\
\rho_e(\phi,\beta_e)&=m_eC_eF_2^+(\phi,\beta_e)\\
n_e(\phi,\beta_e)&=C_eF_2^-(\phi,\beta_e) \\ s_e(\phi,\beta_e)&=\frac{k_B C_e}{\rho}\left[F_1(\phi,\beta_e)+F_3(\phi,\beta_e)-\frac{\phi}{\beta_e}F_2^-(\phi,\beta_e)\right],\label{eq:EPEOS}
\end{align}
the latter of which can be found by noting that
\begin{equation}
s=\frac{u+p-\mu n}{\rho}
\end{equation}
where the specific internal energy is
\begin{equation}
u_e(\phi,\beta_e)=m_ec^2 C_e F_3(\phi,\beta_e)
\end{equation}
ignoring the electron and positron rest mass energy. The integrals that appear in these definitions are: 
\begin{align}
F_1(\phi,\beta_e)&=\int_{\varepsilon=\beta_e}^\infty\Gamma\left(\frac{\varepsilon}{\beta_e}\right)D^+(\varepsilon,\phi)\frac{\mathrm{d}\varepsilon}{\beta_e}\\ F_2^+(\phi,\beta_e)&=\int^\infty_{\varepsilon=\beta_e}\Gamma'\left(\frac{\varepsilon}{\beta_e}\right)D^+(\varepsilon,\phi)\frac{\mathrm{d}\varepsilon}{\beta_e}\\
F_2^-(\phi,\beta_e)&=\int^\infty_{\varepsilon=\beta_e}\Gamma'\left(\frac{\varepsilon}{\beta_e}\right)D^-(\varepsilon,\phi,)\frac{\mathrm{d}\varepsilon}{\beta_e}\\ F_3(\phi,\beta_e)&=\int^\infty_{\varepsilon=\beta_e}\varepsilon\Gamma'\left(\frac{\varepsilon}{\beta_e}\right)D^+(\varepsilon,\phi)\frac{\mathrm{d}\varepsilon}{\beta_e^2}
\end{align}
where $\varepsilon \equiv E/k_BT$,
\begin{align}
\Gamma(x)&\equiv \frac13\left(x^2-1\right)^{\frac32},\quad\textrm{and}\\\quad D^\pm(\varepsilon,\phi)&\equiv\frac{1}{e^{\varepsilon-\phi}+1}\pm\frac{1}{e^{\varepsilon+\phi}+1}.
\end{align}
The partial derivatives of equations \eqref{eq:EPEOS} needed to compute the contribution to the equation of state given in equations \eqref{eq:EOS1} and \eqref{eq:EOS2} can be found by taking appropriate derivatives of these functions. The condition of charge neutrality determines the chemical potential. It is imposed by solving
\begin{equation}
n_e=n_{e^-}-n_{e^+}=\left\langle\frac{Z}{A}\right\rangle\rho
\end{equation}
\ie,~by demanding that the electron excess is due to the ionization of the atoms. We take $\langle Z/A\rangle=1/2$ corresponding to fully ionized ${}^{12}C$ and ${}^{16}$O.

\subsection{Additional bosonic particles}

Similar to the electrons and positrons, we calculate the EOS by integrating over the relevant distribution functions, in this case the Bose-Einstein distribution. We define
\begin{align}
C_{\rm dm}&=\frac{1}{\pi^2}\left(\frac{m_{\rm dm} c}{\hbar}\right)^3=\frac{m_{\rm dm}^3}{m_e^3}C_e,\quad\textrm{and}\\ \beta_{\rm dm}(T)&=\frac{m_{\rm dm} c^2}{k_B T}=\frac{m_{\rm dm}}{m_e}\beta_e.
\end{align}
Note that we are assuming that novel particles are uncharged bosons and, as such, has zero chemical potential. We will denote the spin-degeneracy of the novel particles $X$ by $g_X$ ($g_X=3$ for a hidden photon and $g_X=1$ for an axion). In this case, the thermodynamic quantities are
\begin{align}
P_{\rm dm}(\beta_{\rm dm})&=m_{\rm dm} c^2C_{\rm dm}\left( \frac{g_{\rm dm}}{2}\right)H_1(\beta_{\rm dm}) \\\rho_{\rm dm}(\beta_{\rm dm})&=m_{\rm dm}C_{\rm dm}\left( \frac{g_{\rm dm}}{2}\right)H_2(\beta_{\rm dm})\\
u_{\rm dm}(\beta_{\rm dm})&=m_{\rm dm}c^2C_{\rm dm}\left( \frac{g_{\rm dm}}{2}\right)H_3(\beta_{\rm dm}) \\ s_{\rm dm}&=\frac{k_B C_{\rm dm}}{\rho}\left( \frac{g_{\rm dm}}{2}\right)\left[H_1(\beta_{\rm dm})+H_3(\beta_{\rm dm})\right],\label{eq:DMEOS}
\end{align}
where 
\begin{align}
H_1(\beta_{\rm dm})&=\int_{\varepsilon=\beta_{\rm dm}}^\infty\Gamma\left(\frac{\varepsilon}{\beta_{\rm dm}}\right)B(\varepsilon)\frac{\mathrm{d}\varepsilon}{\beta_{\rm dm}}\\ H_2(\beta_{\rm dm})&=\int^\infty_{\varepsilon=\beta_{\rm dm}}\Gamma'\left(\frac{\varepsilon}{\beta_{\rm dm}}\right) B(\varepsilon)\frac{\mathrm{d}\varepsilon}{\beta_{\rm dm}}\\
H_3(\beta_{\rm dm})&=\int^\infty_{\varepsilon=\beta_{\rm dm}}\varepsilon\Gamma'\left(\frac{\varepsilon}{\beta_{\rm dm}}\right)B(\varepsilon)\frac{\mathrm{d}\varepsilon}{\beta_{\rm dm}^2}\\ B(\varepsilon) &=\frac{1}{e^{\varepsilon}-1}.
\end{align}
One can then take partial derivatives of these functions in order to calculate the contribution of $X$ to equation \eqref{eq:EOS1}.

\bibliography{refs}

\end{document}

%% file: universalnewcommands.tex
\newcommand{\acro}[1]{\textsc{\MakeLowercase{#1}}} %FOR `NOT SHOUTING' CAPS (e.g. acronyms)

\newcommand{\beq}{\begin{equation}}
\newcommand{\eeq}{\end{equation}}
\newcommand{\bea}{\begin{eqnarray}}
\newcommand{\eea}{\end{eqnarray}}

     \newcommand{\cL}{{\cal L}}   \newcommand{\cO}{{\cal O}}  \newcommand{\cQ}{{ \cal Q}}  
\newcommand{\ep}{\epsilon} 
\newcommand{\ie}{{\it i.e.}}  \newcommand{\eg}{{\it e.g.}}
   
\newcommand{\pL}{\left(} \newcommand{\pR}{\right)}      
\newcommand{\alg}[1]{\begin{align} \begin{split} #1 \end{split}  \end{align}}

\newcommand{\refjnl}[1]{{\rm#1}}

\def\apj{\refjnl{ApJ}}                 % Astrophysical Journal
\def\apjs{\refjnl{ApJS}}               % Astrophysical Journal, Supplement
\def\aap{\refjnl{A\&A}}                % Astronomy and Astrophysics